\documentstyle[12pt,equation]{article}
\begin{document}
\newcommand{\tcr}{T_{cr}}
\newcommand{\chit}{\tilde{\chi}}
\newcommand{\phit}{\tilde{\phi}}
\newcommand{\df}{\delta \phi}
\newcommand{\dr}{\delta \rho}
\newcommand{\dkl}{\delta \kappa_{\Lambda}}
\newcommand{\dkg}{\delta \kappa_{G}}
\newcommand{\dkcr}{\delta \kappa_{cr}}
\newcommand{\dxg}{\delta x_{G}}
\newcommand{\dx}{\delta x}
\newcommand{\lx}{\lambda}
\newcommand{\Lx}{\Lambda}
\newcommand{\ex}{\epsilon}
\newcommand{\ks}{k_s}
\newcommand{\gb}{\bar{g}}
\newcommand{\lb}{{\bar{\lambda}}}
\newcommand{\lbz}{\bar{\lambda}_0}
\newcommand{\lt}{\tilde{\lambda}}
\newcommand{\lr}{{\lambda}_R}
\newcommand{\xr}{x_R}
\newcommand{\xt}{\tilde{x}}
\newcommand{\lrt}{{\lambda}_R(T)}
\newcommand{\lbr}{{\bar{\lambda}}_R}
\newcommand{\lk}{{\lambda}(k)}
\newcommand{\lbk}{{\bar{\lambda}}(k)}
\newcommand{\lbkt}{{\bar{\lambda}}(k,T)}
\newcommand{\ltk}{\tilde{\lambda}(k)}
\newcommand{\mx}{{m}^2}
\newcommand{\mxk}{{m}^2(k)}
\newcommand{\mxkt}{{m}^2(k,T)}
\newcommand{\mb}{{\bar{m}}^2}
\newcommand{\mbb}{{\bar{m}}_2^2}
\newcommand{\mbu}{{\bar{\mu}}^2}
\newcommand{\mt}{\tilde{m}^2}
\newcommand{\mr}{{m}^2_R}
\newcommand{\mrt}{{m}^2_R(T)}
\newcommand{\mk}{{m}^2(k)}
\newcommand{\rhoa}{\rho_1}
\newcommand{\rhob}{\rho_2}
\newcommand{\rhb}{\bar{\rho}}
\newcommand{\rhza}{\rho_{10}}
\newcommand{\rhzb}{\rho_{20}}
\newcommand{\rht}{\tilde{\rho}}
\newcommand{\rhz}{\rho_0}
\newcommand{\rhzt}{\rho_0(T)}
\newcommand{\rhztil}{\tilde{\rho}_0}
\newcommand{\rhzk}{\rho_0(k)}
\newcommand{\rhzkt}{\rho_0(k,T)}
\newcommand{\kx}{\kappa}
\newcommand{\kt}{\tilde{\kappa}}
\newcommand{\kk}{\kappa(k)}
\newcommand{\ktk}{\tilde{\kappa}(k)}
\newcommand{\Gammat}{\tilde{\Gamma}}
\newcommand{\Gammak}{\Gamma_k}
\newcommand{\wt}{\tilde{w}}
\newcommand{\be}{\begin{equation}}
\newcommand{\ee}{\end{equation}}
\newcommand{\een}{\end{subequations}}
\newcommand{\ben}{\begin{subequations}}
\newcommand{\beq}{\begin{eqalignno}}
\newcommand{\eeq}{\end{eqalignno}}
\pagestyle{empty}
\noindent
OUTP-95-02 P \\
HD-THEP-94-28 \\
\vspace{3cm}
\begin{center}
{{ \Large  \bf
High Temperature Phase Transition \\
in Two-Scalar Theories
}}\\
\vspace{10mm}
S. Bornholdt,$^{\rm a,}$
\footnote{
Present address:
Institut f\"ur Theoretische Physik, Universit\"at Kiel,
Olshausenstr. 6, 24118 Kiel, Germany.}
N. Tetradis$^{\rm b}$ and C. Wetterich$^{\rm a}$
\vspace {0.5cm}
 \\
a)  {\em
Institut f\"ur Theoretische Physik, Universit\"at Heidelberg,\\
Philosophenweg 16, 69120 Heidelberg, Germany} \\
b) {\em
Theoretical Physics, University of Oxford, \\
1 Keble Road, Oxford OX1 3NP, U.K.}

\end{center}

\setlength{\baselineskip}{20pt}
\setlength{\textwidth}{13cm}

\vspace{3.cm}
\begin{abstract}
{
Two-scalar theories at high temperature exhibit a rich
spectrum of possible critical behaviour, with a
second or first order phase transition.
In the vicinity of the critical temperature
one can observe critical exponents, tricritical points
and crossover behaviour.
None of these phenomena are visible to high temperature
perturbation theory.
}
\end{abstract}
\clearpage
\setlength{\baselineskip}{15pt}
\setlength{\textwidth}{16cm}
\pagestyle{plain}
\setcounter{page}{1}

\newpage

\setcounter{equation}{0}
\renewcommand{\theequation}{{\bf 1.}\arabic{equation}}

\section*{1.
Introduction}

Scalar field theories have been the prototype for
investigations concerning the question of symmetry
restoration at high temperature. Following the
original argument of Kirzhnits and Linde \cite{orig},
the $O(N)$-symmetric scalar theory was considered in
subsequent studies of the problem
\cite{doljack,weinberg,linde}.
The framework in which these studies were carried out
is the perturbative evaluation of the effective
potential \cite{colwein} and its generalization for
non-zero temperature. Even though the restoration
of the spontaneously broken symmetry was qualitatively
demonstrated, the investigation of the
details of the phase transition was not possible,
due to infrared divergences rendering the perturbative
approach unreliable near the critical temperature
\cite{weinberg,linde}.
These divergences originate in the absence of
an infrared cutoff in higher loop contributions
when the temperature dependent mass of the scalar fluctuations
approaches zero near the critical temperature.
An amelioration of the situation was
achieved through the summation of an infinite
subclass of perturbative contributions
(the ``daisy'' graphs) \cite{doljack}. Indeed, these
contributions become dominant for large $N$
and a quantitative description of the phase transition
can be obtained in this limit.
However, the physical picture remained unclear for
small, physically relevant values of $N$, for which
even the order of the transition was not established.
The question was resolved \cite{transition}
through the method of the effective average action
\cite{christof1} - \cite{indices},
which relies on the renormalization group approach.
The phase transition was shown to be second order
for all values of $N$. The quantitative behaviour
near the critical temperature was studied in detail and
the critical system was found to have an effectively
three-dimensional character. Its behaviour can be
characterized by critical exponents
\cite{transition,indices}, in agreement
with known results from three-dimensional field theory.
The picture was verified through an independent
analysis in the large $N$ limit \cite{largen},
with use of other non-perturbative methods, such
as the solution of the Schwinger-Dyson equations.
A summary of the results can be found in ref. \cite{review}.

In a cosmological context first order phase transitions are
more spectacular than second order transitions due to the
departure from thermal equilibrium. One would like to have
a prototype model for a first order transition, for which the
methods of high temperature field theory can be tested, similarly
to the $O(N)$ scalar model for second order transitions.
In statistical physics it is well known that scalar models
with more than one field and discrete symmetries instead of
maximal $O(N)$ symmetry exhibit a rich spectrum of critical
behaviour, including first and second order transitions and
tricritical behaviour in between. Since high temperature
field theories are in close correspondence to
(three-dimensional) statistical models, it seems natural
to investigate such models also as prototypes for first
order transitions in high temperature field theory.
In this paper we apply the method
of the effective average action to the study of
the high temperature phase transitions in theories with
two real scalar fields. The symmetry is not
$O(2)$, but rather a discrete symmetry. This model can serve
as a prototype for a first order phase transition in field theories.
It can be easily generalized to models in which each scalar field
is an $N$-component vector.

We are interested in
the phenomenon of spontaneous symmetry breaking and symmetry
restoration at high temperature. For sufficiently low temperature
our two scalar theory models the Higgs mechanism
in gauge theories,
through which the expectation value
of a scalar field results in a mass term for gauge fields.
\footnote{For sufficiently small gauge coupling the present
investigation and its generalization to the case where each scalar
field has $N$ components gives a reasonable approximation to the
gauged models even in the vicinity of the critical temperature.
However, the determination of the meaning of ``sufficiently small''
needs a detailed investigation of high temperature gauge
theories.}
Perturbative arguments predict a first order phase transition
for this case \cite{first}.
However, the reliability of such predictions is questionable
when the transition becomes weakly first order, due to
infrared divergences similar to the ones plaguing the
study of the $O(N)$-symmetric scalar theory \cite{gaugediv}.
The approximate vanishing of some mass near the critical
temperature results
in the absence of an infrared cutoff
in higher loop contributions of perturbation theory.
In two-scalar theories this is connected with the fact
that one of the fields gets its mass (or part of it)
through the expectation value of the other.

The present work obtains control over
these infrared problems.
\footnote{We should point out that, for a non-abelian
Higgs model,  the situation is much more involved than
the perturbative results indicate, due to the presence
of a confining regime in the symmetric phase of the
model. As this work deals only with scalar fields,
such a complication does not arise. For a discussion
of gauge theories in the context of the effective average
action approach see refs. \cite{gauge,daniel,qcd}.}
Depending on the couplings of the model we find that the
phase transition is either first or second order. For a
sufficiently strongly first order transition
high temperature perturbation theory may give realistic
results for the two-scalar model. We concentrate here
on the more problematic regions of a second order transition,
a weakly first order transition and the tricritical
behaviour at the separation of the two regimes. For the
corresponding values of the couplings high temperature
perturbation theory fails near the critical temperature.
As a byproduct, our results can be used in order to
establish in which region of parameter space
perturbation theory gives a reasonable approximation
for the description of the phase transition.

Our results are directly relevant for two
specific classes of scenaria in the cosmological context:
The first class concerns multi-Higgs-scalar extensions
of the standard model at non-zero temperature.
The prediction of perturbation
theory for a first order electroweak phase transition
\cite{first},
combined with the existence of baryon number violating
processes at non-zero temperature
within the standard model \cite{shaposh},
has generated much interest in the possibility of
creating the baryon asymmetry of the universe
during the electroweak phase transition \cite{sintra}.
It is not clear, however, if the phase transition in the
pure standard model is sufficiently strongly first order
and if there is sufficient $CP$ violation in order to create
an asymmetry of reasonable size.
This has led to the study of
multi-Higgs-scalar extensions of the standard model,
in which the additional scalar fields can be used
to make the phase transition more strongly first
order or to enhance the sources of $CP$ violation in the
model. Also supersymmetric extensions of the standard model
contain two scalar doublets.
It is not clear whether the perturbative methods
used in ref. \cite{twoscalar} for the calculation
of the scalar field contributions to the effective potential
are reliable for such two-scalar models,
if we take into account the ``warning'' from
the study of the $O(N)$-symmetric theory.
Our work gives a reliable estimate of
the effect of these contributions on the nature of the transition.

The second class of scenaria concerns multi-scalar
models of inflation \cite{inflation2}. In most such studies some
classical potential is employed, which may bear no resemblance
to the effective potential.
Thermal effects are often ignored except for the temperature
dependence
of the mass term. If inflation is initiated by a high temperature
phase transition
our formalism sets the framework for
the proper study of the problem.

We consider a theory of two real scalar fields
$\chi_a (a=1,2)$,
invariant under the discrete symmetries
$(\chi_1 \leftrightarrow - \chi_1,
\chi_2 \leftrightarrow - \chi_2,
\chi_1 \leftrightarrow  \chi_2)$,
which we denote by
$(1 \leftrightarrow - 1,
2 \leftrightarrow - 2,
1 \leftrightarrow  2)$ for brevity.
The symmetry group is $Z_4 \times Z_2$, consisting of
$90^o$ rotations in the ($\chi_1$, $\chi_2$) plane and
a reflection on one of the axes.
The classical potential can be written
as
\beq
V(\chi_1,\chi_2)
=& \frac{1}{2} \mb (\chi_1^2+\chi_2^2)
+ \frac{1}{8} \lb (\chi_1^4+\chi_2^4)
+ \frac{1}{4} \gb \chi_1^2\chi_2^2
\nonumber \\
=& \frac{1}{2} \mb (\chi_1^2+\chi_2^2)
+ \frac{1}{8} \lb (\chi_1^2+\chi_2^2)^2
+ \frac{1}{4} x \lb \chi_1^2\chi_2^2,
\label{oneone} \eeq
with
\be
x=\frac{\gb}{\lb} - 1.
\label{onetwo} \ee
For $V$ to be bounded from below we
require $\lb > 0$, $x > -2$.

For $\mb>0$ the classical theory is in the symmetric regime
(which we denote by S)
with the minimum of the classical potential at the origin.
For $\mb<0$ the theory is in the spontaneously broken regime and
we distinguish two possibilities
consistent with the
symmetry:\\
I) For $x < 0$ four degenerate minima of the potential are located
between the two axes at
\be
\chi_{10}= \pm \chi_{20}= \pm \sqrt{- \frac{2 \mb}{\lb + \gb}}.
\label{hir1}
\ee
We denote this regime
by M. \\
II) For $x>0$ the four
minima of the potential are located on
the axes at
\be
\chi_{10}= \pm \sqrt{- \frac{2 \mb}{\lb}},~~~ \chi_{20}=0,
\label{hir2}
\ee
or similarly with $\chi_{10}$ and $\chi_{20}$ interchanged.
We denote this regime by AX.\\
The regimes M and AX are closely related.
A redefinition of the fields according to
\beq
\chit_1 &= \frac{1}{\sqrt{2}} (\chi_1 + \chi_2)
\nonumber \\
\chit_2 &= \frac{1}{\sqrt{2}} (\chi_1 - \chi_2)
\label{onethree} \eeq
results in a rotation of the axes by $45^o$, thus transforming
the AX into the M regime.
The couplings of the redefined theory are related to the old
ones according to
\beq
\lt &= \lb \left( 1 + \frac{x}{2} \right)
\nonumber \\
\xt &= - \frac{x}{1 + \frac{x}{2}}.
\label{onefour} \eeq

There are three characteristic values of $x$: \\
a) For $x=0$ the symmetry of the theory is increased to
$O(2)$. \\
b) For $x=-1$ the theory decomposes
into two disconnected
$Z_2$-symmetric
models for $\chi_1$ and $\chi_2$ separately. \\
c) Similarly, for $x=2$ (and therefore $\xt=-1$, according
to eqs. (\ref{onefour})) the theory decomposes
into two disconnected
$Z_2$-symmetric
models for $\chit_1$ and $\chit_2$. \\
The above symmetries are expected to be preserved after the
quantum or thermal corrections have been taken into account.
This means that any renormalization group flow of the couplings
that starts on the surfaces $x=0,-1,2$ in parameter space
cannot take the system out of them.
As a result the parameter space of the theory is
divided into the four regions:
$x>2$, $2>x>0$, $0>x>-1$, $x<-1$, which are not connected by
the renormalization group flow of the couplings.
The phase transitions for the theories which correspond
to $x=0,-1,2$ have been discussed in detail
in refs. \cite{transition,indices}. They are second order
transitions governed by effectively three-dimensional fixed
points. In our model these fixed points exist
on surfaces separating the parameter space
into disconnected regions. We shall demonstrate all the
above points in the following sections.

We should point out that
this model was discussed in ref. \cite{pert1}
through use of finite temperature perturbation theory.
No part of the rich structure of critical behaviour
that we shall describe
in the following sections was observed.
The universal, effectively three-dimensional
behaviour of the system near the critical temperature
is common for statistical systems and three-dimensional
field theories which belong to the same universality class.
(The statistical systems are
characterized as two-component spin systems with cubic
anisotropy.)
As a consequence, various aspects of
the problem have been investigated
in refs. \cite{aharony} - \cite{march} (and references therein)
through other methods. Our results are in very good agreement
with all these studies.
Similar models have been considered in ref. \cite{jain}.

The outline of our procedure is the following: We
make use of the effective average action
$\Gamma_k$, which results from the effective integration
of quantum and thermal fluctuations with characteristic
momenta $q^2 > k^2$. It contains all the information
on the generalized couplings of the theory and their
dependence on the scale $k$.
For $k$ of the order of some ultraviolet
cutoff $\Lx$ the effective average action
is equal to the classical (bare) action (no integration
of fluctuations takes place). For $k=0$,
$\Gamma_k$ is equal to the effective action (all fluctuations
are integrated).
The dependence of $\Gamma_k$ on
the scale $k$ is given by an exact non-perturbative
renormalization group
equation, which can be expressed as evolution
equations for the running couplings of the theory.
These equations can be solved
within some appropriate
approximation scheme, with the classical couplings as
initial conditions for $k=\Lx$.
The renormalized couplings of the
theory are
obtained for $k=0$. The calculation can be performed
for zero and non-zero temperature.
The gradual incorporation of the
effects of quantum and thermal fluctuations into the
running couplings is the essential element
which resolves the problem of infrared divergences
that invalidates perturbative schemes.
The basic formalism of the effective average action
is summarized in section 2.
We expect that the running of the couplings
for $k$ much smaller than the temperature has an
effectively three-dimensional character. The
reason is that
the effective dimensionality is reduced,
when the characteristic length scale
$1/k$ of the ``coarse grained'' system is much larger
than the periodicity $1/T$ in the imaginary time direction
set by the temperature.
This is expected to be important near the critical temperature,
where no infrared cutoffs (such as masses)
other than $k$ exist.
As a result, the fixed point structure of the
three-dimensional theory determines the behaviour
of the critical system. For this reason
we present in section 3
a qualitative study of the three-dimensional
theory and its fixed points, based on a crude approximation
scheme for the solution of the exact renormalization
group equation.
In section 4, we develop more elaborate
(and, therefore, more accurate) approximation schemes.
They are generalized for non-zero temperature in section 5.
These tools are put into work in sections 6 and 7: The
evolution of the running couplings is calculated, starting
with the classical theory at
scales $k =\Lx \gg T$ and finishing at $k=0$, where the renormalized
theory is obtained.
In section 7 we explicitly demonstrate how the evolution
of the running couplings becomes effectively three-dimensional
for $k \ll T$. In section 8 we calculate the critical temperature
for the phase transition.
In sections 9-11 we discuss the details of this transition.
We observe a rich spectrum of critical behaviour with critical
exponents, crossover phenomena, tricritical points etc. None
of these are visible within perturbation theory. Our conclusions
are given in section 12.

\setcounter{equation}{0}
\renewcommand{\theequation}{{\bf 2.}\arabic{equation}}

\section*{2. The evolution equation for the effective
average potential}

We consider a theory of two real scalar fields
$\chi_a (a=1,2)$, in
$d$-dimensional Euclidean space, with an
action $S[\chi]$ invariant under the
$(1 \leftrightarrow - 1,
2 \leftrightarrow - 2,
1 \leftrightarrow  2)$
symmetry.
We specify the action together with some ultraviolet
cutoff $\Lambda$, so that the theory is properly
regulated.
We add to the kinetic term an infrared regulating piece \cite{exact}
\be
\Delta S = \frac{1}{2} \int \frac{d^d q}{(2 \pi)^d}
R_k(q) \chi^*_a(q) \chi^a(q),
\label{twoone} \ee
where $\chi^a(q)$ are the Fourier modes of the scalar fields.
The function $R_k$  is employed in
order to prevent the propagation of modes
with characteristic momenta $q^2 < k^2$.
This can be achieved, for example,
by the choice
\be
R_k(q) = \frac{q^2 f^2_k(q)}{1 - f^2_k(q)},
\label{twotwo} \ee
with
\be
f^2_k(q) = \exp \left( - \frac{q^2}{k^2} \right).
\label{twothree} \ee
We point out that there are many alternative choices for $R_k(q)$, some
of which were used in refs.
\cite{christof1} - \cite{review}.
The physical results which are obtained when the cutoff
is removed are scheme independent.
The choice of eqs. (\ref{twotwo}),(\ref{twothree})
is the most natural one \cite{exact} and convenient for
numerical calculations.
For a massless field the inverse propagator
derived from the action $S + \Delta S$ has a minimum $\sim k^2$.
The modes with $q^2 \gg k^2$ are unaffected by the infrared cutoff,
while the
low frequency modes with $q^2 \ll k^2$ are cut off,
as $R_k$ acts like a mass term
\be
\lim_{q^2 \rightarrow 0} R_k(q) = k^2.
\label{twothreepr} \ee
We subsequently introduce sources and
define the generating functional for the connected Green functions
for the action $S + \Delta S$. Through a Legendre
transformation we obtain the
generating functional for the 1PI Green functions
${\tilde \Gamma}_k[\phi^a]$, where $\phi^a$ is the expectation value of the
field $\chi^a$ in the presence of sources.
The use of the modified propagator for the calculation of
${\tilde \Gamma}_k$ results in the effective integration of only the
fluctuations with $q^2 >
k^2$. Finally, the effective average action is
obtained by removing the infrared cutoff
\be
\Gamma_k[\phi^a] = {\tilde \Gamma}_k[\phi^a] -
\frac{1}{2} \int \frac{d^d q}{(2 \pi)^d}
R_k(q) \phi^*_{a}(q) \phi^a(q).
\label{twofour} \ee
For $k$ equal to the ultraviolet cutoff $\Lambda$, $\Gammak$ becomes
equal
to the classical action $S$ (no effective integration of modes takes
place), while for $k \rightarrow 0$ it tends towards the effective action
$\Gamma$ (all the modes are included)
which is the generating functional of the 1PI Green functions computed
from $S$ (without infrared cutoff).
For intermediate values of $k$ the effective average action
realizes the concept of a coarse grained effective action in the
sense of ref. \cite{langer}.

The interpolation of $\Gammak$ between the classical and the
effective action makes it a very useful field theoretical tool.
The means for practical calculations is provided by an exact
flow equation
\footnote{
See ref. \cite{reneq} for other versions of exact renormalization
group equations.}
which describes the response of the
effective average action to variations of the infrared cutoff
($t=\ln (k/\Lambda)$) \cite{exact}
\be
\frac{\partial}{\partial t} \Gammak[\phi]
= \frac{1}{2} {\rm Tr} \bigl\{ (\Gammak^{(2)}[\phi] + R_k)^{-1}
\frac{\partial}{\partial t} R_k \bigr\}.
 \label{twofive} \ee
Here $\Gammak^{(2)}$ is the second functional derivative of the effective
average action with respect to $\phi^a$. For
real fields it reads in momentum space
\be
(\Gammak^{(2)})^a_b(q,q') =
\frac{\delta^2 \Gammak}{\delta \phi^*_a(q) \delta \phi^b(q')},
\label{twosix} \ee
with
\be
\phi^a(-q)=\phi^*_a(q).
\label{twoseven} \ee
The non-perturbative flow equation
has the form of an
one-loop expression
involving the exact inverse propagator $\Gamma^{(2)}_k$ together
with an
infrared cutoff provided by $R_k$.
No contributions from higher loops appear in this exact equation.

For the solution of eq. (\ref{twofive})
one has to develop an efficient truncation scheme.
The form of the effective average action is constrained by the
$(1 \leftrightarrow - 1,
2 \leftrightarrow - 2,
1 \leftrightarrow  2)$
symmetry.
However, there is still an infinite number of invariants
to be considered.
Throughout this paper we shall work with
an approximation which neglects the
effects of wave function renormalization. We shall, therefore,
keep only a classical kinetic term in the effective average
action
\be
\Gammak =
\int d^dx \bigl\{ U_k(\rhoa, \rhob)
+ \frac{1}{2} \partial^{\mu} \phi_a
\partial_{\mu} \phi^a \bigr\},
\label{twoeight} \ee
and neglect all invariants which involve more derivatives
of the fields.
We have used the definition
$\rho_1 = \frac{1}{2} \phi_1^2$ and similarly for
$\rho_2$.
The justification for our approximation lies in the
smallness of the anomalous dimension,
which is expected to be $\eta \simeq 0.03-0.04$
for the three-dimensional theory.
We estimate the corrections arising from the proper
inclusion of wave function renormalization effects to
be of the same order as $\eta$ (a few \%).
An improved treatment will be given elsewhere \cite{peter}.
In order to obtain an evolution equation for $U_k$
from eq. (\ref{twofive}),
we have to expand around a constant
field configuration (so that the derivative terms in the parametrization
(\ref{twoeight}) do not contribute to the l.h.s. of eq. (\ref{twofive})).
Eq. (\ref{twofive}) then gives \cite{christof2} - \cite{indices}
\be
\frac{\partial}{\partial t} U_k(\rhoa,\rhob) =
\frac{1}{2} \int \frac{d^d q}{(2 \pi)^d}
\left( \frac{1}{P(q^2) + M^2_1}
+\frac{1}{P(q^2) + M^2_2} \right)
\frac{\partial}{\partial t} R_k(q).
\label{twonine} \ee
$P(q^2)$ results from
the combination of
the classical kinetic contribution $q^2$
and the regulating term $R_k$
into an effective
inverse propagator (for massless fields)
\be
P(q^2) = q^2 + R_k = \frac{q^2}{1 - f^2_k(q)},
\label{twothirteen} \ee
with $f^2_k(q)$ given by eq. (\ref{twothree}).
For $q^2 \gg k^2$ the inverse ``average'' propagator $P(q)$ approaches
the standard inverse propagator $q^2$ exponentially fast, whereas
for $q^2 \ll k^2$ the infrared cutoff
prevents the propagation.
$M^2_{1,2}$ are the eigenvalues of the mass matrix at the
point $(\rhoa,\rhob)$
\beq
M^2_{1,2}(\rhoa,\rhob) =
\frac{1}{2} \biggl\{
&U_1 + U_2 +2 U_{11} \rhoa + 2 U_{22} \rhob
\nonumber \\
&\pm \left[
(U_1 - U_2 +2 U_{11} \rhoa - 2 U_{22} \rhob )^2
+ 16 U^2_{12} \rhoa \rhob \right]^{\frac{1}{2}}
\biggr\},
\label{twoten} \eeq
and we have introduced the
notation $U_1 = \frac{\partial U_k}{\partial \rhoa}$,
$U_{12} = \frac{\partial^2 U_k}{\partial \rhoa \partial \rhob}$ etc.

Eq. (\ref{twonine}) is the master equation for our investigation.
It is a non-linear partial differential equation
for three independent
variables
$(t, \rhoa, \rhob)$. Since it is difficult to solve it
exactly we again resort to some approximation scheme.
We first introduce a Taylor expansion of
$U_k(\rhoa,\rhob)$ around its minimum.
This turns eq. (\ref{twonine}) into an infinite system of
ordinary (coupled) differential equations for
the $k$-dependence of the
minimum and the derivatives of the effective average
potential, with
independent variable $t=\ln (k/\Lambda)$.
We solve this system approximately
by truncating at a finite number of derivatives.
This approach has been used in the past for the study of
the $O(N)$-symmetric scalar theory.
It has provided a full, detailed picture of the
high temperature phase transition for this theory
\cite{transition},
\cite{indices} - \cite{review}, with accurate determination
(at the few \% level) of
such non-trivial quantities as the critical
exponents \cite{indices}.
An estimate \cite{tim1} of the residual errors for high
level truncations indicates that
they are smaller than the uncertainties introduced
by the imprecise treatment of the wave function
renormalization effects.
For this work we shall use the lowest level
truncation which keeps only the second derivatives
of the potential $U_{11},U_{22},U_{12}$. This will be
sufficient for a reliable determination of the phase
diagram and a crude estimate of universal quantities such
as critical exponents and crossover curves.
For an improved treatment
see ref. \cite{stefan}, and for a discussion which
takes into account the next level in the truncation for $U_k$
and the first corrections arising from
wave function renormalization see ref. \cite{peter}.

\setcounter{equation}{0}
\renewcommand{\theequation}{{\bf 3.}\arabic{equation}}

\section*{3.
The phase structure of the three-dimensional theory}

Before performing a more detailed analysis we would like
to gain some understanding of the phase structure of the
theory. As we have already mentioned in the introduction,
the behaviour of the four-dimensional theory
near a high temperature second order phase transition
is expected to have a three-dimensional character.
The reason for this is the divergence of the correlation
length for the fluctuations of the system (the mass of
some field goes to zero). As a result, the characteristic
length scale for the critical system is much larger than
the periodicity in the imaginary time
direction due to temperature (for details see the following
sections).
\footnote{If the phase transition is strongly first order
this need not be true, because the mass of the fluctuations
does not go to zero and the correlation length does not
diverge near the transition.}
This leads to dimensional reduction and
the critical system has effectively three-dimensional
behaviour. For this reason we are interested in the
phase structure of the three-dimensional theory. More specifically
we want to investigate the possible existence of fixed points
which govern the dynamics of second order phase transitions.

For the purpose of this section we parametrize the potential
by its derivatives at the origin (S regime)
\beq
\mb(k) &= U_1(0) = U_2(0)  \nonumber \\
\lb(k) &= U_{11}(0) = U_{22}(0)  \nonumber \\
\gb(k) &= U_{12}(0) \nonumber \\
x(k)   &= \frac{\gb(k)}{\lb(k)} -1.
\label{lala} \eeq
The equality of $U_1$, $U_2$ and
$U_{11}$, $U_{22}$ is imposed  by the
$(1 \leftrightarrow - 1,
2 \leftrightarrow - 2,
1 \leftrightarrow  2)$
symmetry of the theory.
For the potential to be bounded we
also require $x > -2$.
For a rough estimate the three-dimensional couplings
are related to the effective couplings of the four-dimensional
theory at high temperature by
$\lb(2 \pi T) = \lx_4 T$,
$\gb(2 \pi T) = g_4 T$,
$\mb(2 \pi T) = m^2_4 + c T^2$,
with appropriate $c$ (for details see sections 6 and 7).
Evolution equations for the above parameters are obtained
by taking derivatives of eq. (\ref{twonine}) with respect
to $\rho_{1,2}$.
The truncated flow equations read for $d=3$
\beq
\frac{d {\bar m}^2}{d t} &= v_3 k  (4+x)  \lb L_1^3({\bar m}^2)
\label{fiveone} \\
\frac{d \lb}{d t} &= - v_3 k^{-1} (10+2x+x^2)
\lb^2 L_2^3({\bar m}^2)
\label{fivetwo} \\
\frac{d x}{d t} &= v_3 k^{-1} (x+1)x(x-2)
\lb  L_2^3({\bar m}^2),
\label{fivethree}
\eeq
with $v_3 = 1/8\pi^2$.
The threshold functions
$L^3_n(w)$
suppress the contributions of massive modes to the evolution equations.
They are
studied in detail in the following sections and in appendix A.
It is convenient to define the dimensionless couplings
\beq
m^2(k) =~& \frac{\mb(k)}{k^2}
\nonumber \\
\lx(k) =~& \frac{\lb(k)}{k}
\nonumber \\
g(k) =~& \frac{\gb(k)}{k}.
\label{fivefive} \eeq
In terms of these quantities the evolution equations have a
scale invariant form, in the sense that the r.h.s. does not
explicitly involve a dependence on $k$
\beq
\frac{d m^2}{d t} &= -2 m^2 + v_3   (4+x)  \lx L_1^3(m^2)
\label{fivesix} \\
\frac{d \lx}{d t} &= -\lx - v_3  (10+2x+x^2)
\lx^2 L_2^3(m^2)
\label{fiveseven} \\
\frac{d x}{d t} &= v_3 (x+1)x(x-2)
\lx  L_2^3(m^2).
\label{fiveeight}
\eeq

We are interested in the
fixed points of the last set of
equations.
For any $x$,
there is an ultraviolet attractive Gaussian fixed point
with $m^2=\lx=0$.
There are also three fixed points with at least one
infrared attractive direction \cite{aharony} - \cite{march}.
They all appear for $m^2 <0$, $\lx >0$. (The exact values are
not important since the discussion in this section is only
qualitatively correct.)
For their identification we use their standard names
in statistical physics \cite{aharony,amit}.    \\
a) The Heisenberg fixed point has $x=0$ and corresponds to
a theory with symmetry increased to $O(2)$, as we have
discussed in the introduction. \\
b) The Ising fixed point has $x=-1$ and corresponds to two
disconnected $Z_2$-symmetric theories. \\
c) The Cubic fixed point has $x=2$ and corresponds to two
disconnected theories, if the fields are redefined
according to eqs. (\ref{onethree}). \\
All these points are infrared unstable in the $m^2$ direction
and are located on a critical surface
$m^2_{cr} = m^2_{cr}(\lx,x) < 0$. Solutions of the
evolution equations which start above the critical surface,
with $m^2 > m^2_{cr}$, flow towards the region of positive
$m^2$ for $t \rightarrow -\infty$,
and correspond to theories in the symmetric phase.
Solutions with
$m^2 < m^2_{cr}$ flow deep into the region of negative $m^2$
and correspond to theories in the phase
with spontaneous symmetry breaking.

The relative stability of the fixed points on the critical
surface determines which one governs the dynamics of the
phase transition very close to the critical
temperature. For
a first simple investigation of the relative stability
in the ($\lx$, $x$) directions we fix $m^2$ to an arbitrary value
(we choose $m^2 = 0$ for convenience)
and solve eqs. (\ref{fiveseven}), (\ref{fiveeight}) numerically.
The results are presented in fig. 1. We observe that all three
fixed points are attractive in the $\lx$ direction. However, the
Ising and Cubic fixed points are repulsive in the $x$ direction,
while the Heisenberg fixed point is totally attractive.
We observe four disconnected regions: \\
a) $2 >x>0$ : The trajectories flow away from the Cubic
and towards the Heisenberg fixed point. \\
b) $0> x >-1$ : The trajectories flow away from the Ising
and towards the Heisenberg fixed point. \\
c) $x >2$ : The trajectories flow
away from the Cubic fixed point and into a region of large $x$ and
small $\lx$. Eventually $\lx$ turns negative at a finite value
of $k$. (This can be verified through the explicit solution
of eqs. (\ref{fiveseven}), (\ref{fiveeight}) in this region.)
At this point an instability arises, as the potential seems not
to be bounded from below. Our treatment is not sufficient for a
detailed investigation of the nature of this instability, since
our truncation scheme is very crude.
A detailed discussion is given in ref. \cite{stefan}, where improved
truncations are employed. It is shown that the instability is
not real since the higher derivatives of the potential
remain positive. The change of sign for $\lx$ corresponds to the
disappearance of a false vacuum of the theory and results in
a first order phase transition. We shall return to this point
in the following sections. \\
d) $x < -1$ : The trajectories flow
away from the Ising fixed point and
cross the line $x=-2$ at a finite $k$.
This again implies the presence of an instability whose
true nature is related to the disappearance of a false vacuum.
The model exhibits a first order transition also for $x < -1$.\\
Flows that start on the lines $x=0,-1,2$ remain on these lines.
No trajectories exist which connect the four regions
$x>2$, $2>x>0$, $0>x>-1$, $x<-1$.
All this is in agreement with the discussion at the end of the
introduction.

The diagram of fig. 1 determines the phase structure of the
theory when
the behaviour of the system
becomes effectively three-dimensional (i.e. close to the
critical temperature).
For parameters in the regions
$2 >x>0$, $0> x >-1$ we expect second order phase transitions,
with critical dynamics governed by the three fixed points.
These two regions can be mapped onto each other
through a redefinition of the fields according to
eqs. (\ref{onethree}), (\ref{onefour}). This indicates that
the Ising and Cubic fixed point should lead to
identical universal behaviour (and
therefore to identical universal quantities, such as critical
exponents).
Very close to the critical temperature we expect the
Heisenberg fixed point to dominate the dynamics, but the
other two can be relevant if the initial values of the
running parameters are sufficiently close to them.
In the parameter regions
$x>2$, $x<-1$ we expect first order phase transitions.
In the following sections
we shall verify the above conclusions with improved
quantitative accuracy.

\setcounter{equation}{0}
\renewcommand{\theequation}{{\bf 4.}\arabic{equation}}

\section*{4. Truncations of the evolution equation}

We proceed now to a more detailed study
of the evolution equation and its truncations.
As we have discussed at the end of section 2,
we parametrize the effective average potential
by its minimum and its derivatives at the minimum.
For this work we shall use a truncation that preserves
up to second derivatives of the potential.

In the {\it symmetric regime} (which we denote by S)
the minimum of the
potential is at $\rhza(k)=\rhzb(k)=0$
and we use the definitions of eqs. (\ref{lala}).
The evolution equations in arbitrary dimension $d$ preserve
automatically the symmetry. They read
\beq
\frac{d \mb}{d t} &= v_d k^{d-2} (4+x)  \lb L_1^d(\mb)
\label{threetwo} \\
\frac{d \lb}{d t} &= - v_d k^{d-4} (10+2x+x^2)
\lb^2 L_2^d(\mb)
\label{threethree} \\
\frac{d x}{d t} &= v_d k^{d-4} (x+1)x(x-2)
\lb  L_2^d(\mb),
\label{threefour}
\eeq
with the dimensionless integrals $L^d_n(w)$ given by
\beq
L^d_n(w) = &
- n k^{2n-d} \pi^{-\frac{d}{2}} \Gamma \left( \frac{d}{2} \right)
\int d^d q \frac{\partial P}{ \partial t} (P + w)^{-(n+1)} \nonumber \\
= &
- n k^{2n-d}
\int_0^{\infty} dx x^{\frac{d}{2}-1}
\frac{\partial P}{ \partial t} (P + w)^{-(n+1)}.
\label{threefive} \eeq
Here $P$ is given by eq. (\ref{twothirteen}), and
\be
v_d^{-1} = 2^{d+1} \pi^{\frac{d}{2}} \Gamma
\left( \frac{d}{2} \right).
\label{threevd} \ee

In the {\it spontaneously broken regime} there are
two possibilities consistent with the
symmetry:

I) In the M regime
the minimum of the potential is located
symmetrically between the $\rho$-axes at
$\rhza(k)=\rhzb(k)=\frac{1}{2} \rhz(k)$.
We define the couplings
\beq
\lb(k) &= U_{11}(\rhz) = U_{22}(\rhz) > 0 \nonumber \\
\gb(k) &= U_{12}(\rhz) \nonumber \\
x(k)   &= \frac{\gb(k)}{\lb(k)} -1.
\label{threesix} \eeq
The requirement that the point
$(\frac{1}{2} \rhz,\frac{1}{2} \rhz)$
is the minimum of the potential imposes $x < 0$, while the
potential is bounded at infinity for $x > -2$.
For $x=0$ the symmetry of the theory is increased to
$O(2)$ and the potential develops
a series of degenerate minima along the circle
$\rhza+\rhzb = \rhz$. For $x=-1$ the theory decomposes
into two disconnected
$Z_2 (\phi_{1,2} \leftrightarrow - \phi_{1,2})$-symmetric
models.
The mass eigenvalues are given by
$M^2_1 = (2+x) \lb \rhz$,
$M^2_2 = - x \lb \rhz$.
The evolution equation for the minimum $\rhz(k)$ is obtained by
considering the total $t$-derivative
of the conditions
$\frac{\partial U_k}{\partial \rhoa}|_{\rhz} =
\frac{\partial U_k}{\partial \rhob}|_{\rhz} = 0$
\cite{transition,christof2}. Again, the truncated evolution equations
automatically preserve the
symmetry and read
\beq
\frac{d \rhz}{d t} =~&- v_d k^{d-2}   \bigl\{
3 L_1^d((2+x) \lb \rhz)
+ \frac{2-x}{2+x}  L_1^d(-x \lb \rhz) \bigr\}
\label{threeseven}
\\
\frac{d \lb}{d t} =~&- v_d k^{d-4}
\frac{3 x \lb}{\rhz} \frac{1+\frac{x}{4}}{1+x}
\bigl\{ L_1^d((2+x)\lb \rhz )
- L_1^d(-x \lb \rhz) \bigr\}
\nonumber \\
&- v_d k^{d-4} \lb^2
\biggl\{ 9 \left( 1+\frac{x}{2} \right)^2
L_2^d((2+x)\lb \rhz) +
\left( 1-\frac{x}{2} \right)^2
L_2^d(-x\lb \rhz) \biggr\}
\label{threeeight}
\\
\frac{d x}{d t} =~& v_d k^{d-2}
\frac{3}{\rhz} \frac{2+x}{1+x} \left( x+\frac{x^2}{4} \right)
\bigl\{ L_1^d((2+x)\lb \rhz) - L_1^d(-x \lb \rhz) \bigr\}
\nonumber \\
&+ v_d k^{d-4} x \lb
\biggl\{
9 \left( 1+\frac{x}{2} \right)^2 L_2^d((2+x)\lb \rhz)
+ \left( 1-\frac{x}{2} \right)^2  L_2^d(-x \lb \rhz) \biggr\}.
\label{threenine}
\eeq
For $x=0$ the above evolution equations reproduce the
equations of the $O(2)$-symmetric theory, while for
$x=-1$ those of the $Z_2$-symmetric one
(compare with refs. \cite{transition,indices}).

II) In the regime which we denote by AX,
two degenerate minima of the potential exist on each one of the
$\rho$-axes.
Without loss of generality we concentrate on the minimum at
$\rhza(k)=\rhz(k), \rhzb(k)=0$.
At the level of truncations that we are considering,
the remaining parameters of the theory are conveniently
defined as
\beq
\lb(k) &= U_{11}(\rhz)  \nonumber \\
\gb(k) &= U_{12}(\rhz) \nonumber \\
x(k)   &= \frac{\gb(k)}{\lb(k)} -1 \nonumber \\
\mbb(k) &= U_2(\rhz).
\label{threeten} \eeq
The symmetry demands that
for the truncated potential
\be
\mbb(k) = x(k) \lb(k) \rhz(k).
\label{threxx} \ee
The requirement that the point
$(\rhz,0)$
is the minimum of the potential imposes $x > 0$.
As before, for $x=0$ the symmetry of the theory is increased to
$O(2)$.
The mass eigenvalues are given by
$M^2_1 = 2 \lb \rhz$,
$M^2_2 =  x \lb \rhz$.
At this point we encounter a difficulty.
The derivation of truncated evolution equations
is hindered by the fact that
the parametrization around a minimum located on
one of the axes is asymmetric between the two fields.
As a result the symmetry
$(\phi_1 \leftrightarrow \phi_2)$, is not
maintained by the evolution equations
at each level of the truncations.
More specifically, the
flow equations for the
couplings $U_{11}(\rhz)$, $U_{22}(\rhz)$ are different.
Also eq. (\ref{threxx})
is not preserved by the evolution equation.
This is not
surprising, since these relations are not expected to hold
for the exact potential without truncation.
It is easy to see that
they are altered as soon as third derivatives of the potential
are included.
This is in constrast with what
happens in the M regime, where
the formulation is symmetric between the two fields.
For example, in the M regime
the couplings $U_{11}$, $U_{22}$ are expected
to remain equal at every level of truncations, and indeed this
is guaranteed by the evolution equations.
The above remarks indicate a natural
method of preserving the
$(1 \leftrightarrow - 1,
2 \leftrightarrow - 2,
1 \leftrightarrow  2)$
symmetry in the AX regime.
A redefinition of the fields according to
\beq
\phit_1 &= \frac{1}{\sqrt{2}} (\phi_1 + \phi_2)
\nonumber \\
\phit_2 &= \frac{1}{\sqrt{2}} (\phi_1 - \phi_2)
\label{threeeleven} \eeq
results in a rotation of the axes by $45^o$, thus transforming
the AX into the M regime.
The couplings of the redefined theory are related to the old
ones according to
\beq
\rhztil &= \rhz \nonumber \\
\lt &= \lb \left( 1 + \frac{x}{2} \right)
\nonumber \\
\xt &= - \frac{x}{1 + \frac{x}{2}}.
\label{threetwelve} \eeq
The evolution equations for the new quantities have
already been worked out and are given
by eqs. (\ref{threeseven})-(\ref{threenine}).
By simply rewriting them in terms of the old
quantities defined in eqs. (\ref{threeten}) we obtain
\beq
\frac{d \rhz}{d t} =~& - v_d k^{d-2}
\bigl\{ 3 L_1^d(2\lb \rhz) + (1+x)  L_1^d(x \lb \rhz) \bigr\}
\label{threethirteen}
\\
\frac{d \lb}{d t} =~& -v_d k^{d-4}
\lb^2
\bigl\{ 9 L_2^d(2\lb \rhz) + (1+x)^2  L_2^d(x \lb \rhz) \bigr\}
\label{threefourteen}
\\
\frac{d x}{d t} =~& v_d k^{d-2}
\frac{6}{\rhz}  \frac{x+\frac{x^2}{4}}{1-\frac{x}{2}}
\bigl\{ L_1^d(2\lb \rhz) - L_1^d(x \lb \rhz) \bigr\}
\nonumber \\
&+ v_d k^{d-4} x \lb
\bigl\{
9 L_2^d(2 \lb \rhz)
+ (1+x)^2
L_2^d(x \lb \rhz) \bigr\}.
\label{threefifteen}
\eeq
For $x=0$ the above evolution equations reproduce the ones of
the $O(2)$-symmetric theory. Another special point is
$x=2$ for which the theory, when
expressed in terms of the redefined fields $\phit_1$, $\phit_2$,
decomposes into two disconnected
$Z_2 (\phit_{1,2} \leftrightarrow - \phit_{1,2})$-symmetric
models.
We should point out that eqs.
(\ref{threethirteen})-(\ref{threefifteen})
could have been obtained by defining
$\lb=U_{11}(\rhz)$ and $x=U_2(\rhz)/U_{11}(\rhz)\rhz$
(in agreement with eq. (\ref{threxx})) and
inserting eqs. (\ref{threeten})
on the r.h.s. of the flow equations.
The advantage of the
redefinition (\ref{threeeleven}) is that it makes transparent
how this apparently arbitrary choice of parameters
preserves the original
symmetry at this truncation level.

In the following sections we shall use the evolution equations
(\ref{threetwo})-(\ref{threefour}),
(\ref{threeseven})-(\ref{threenine}),
(\ref{threethirteen})-(\ref{threefifteen}),
for the S, M, AX regimes respectively, in order to
obtain the renormalized theory in its various phases.

\setcounter{equation}{0}
\renewcommand{\theequation}{{\bf 5.}\arabic{equation}}

\section*{5. The integrals $L^d_n$ for zero and non-zero
temperature}

The integrals $L^d_n(w)$, defined in eq. (\ref{threefive}),
have been discussed extensively in refs.
\cite{christof2,indices,convex}
(for various shapes of the infrared regulating function $R_k(q)$,
for which eqs. (\ref{twotwo}),(\ref{twothree}) are the most
natural choice \cite{exact}).
For completeness, we give in appendix A a summary of the
properties of $L^d_n(w)$. Also,
as an example, we plot in fig. 2 the integrals $L^3_1(w)$,  $L^3_2(w)$.
Their most interesting property, for our discussion, is that
they fall off for large values of $w/k^2$, following a power law. As
a result they introduce threshold behaviour for the
contributions of massive modes to the evolution equations.
It is obvious from
eqs. (\ref{threetwo})-(\ref{threefour}),
(\ref{threeseven})-(\ref{threenine}),
(\ref{threethirteen})-(\ref{threefifteen}),
for the S, M, AX regimes respectively,
that the various contributions to the evolution equations
involve $L^d_n$ integrals with the mass eigenvalues as
their arguments.
When the running squared
mass of a massive mode
becomes much larger than the scale $k^2$
(at which the system is probed), these
contributions vanish and the massive modes decouple.
We evaluate the integrals $L^d_n(w)$
numerically and use numerical fits for the solution of
the evolution equations.

In order to extend the formalism of the
previous section to non-zero temperature
we only need to recall that,
in Euclidean formalism, non-zero temperature $T$ results
in periodic boundary conditions in the time direction
(for bosonic fields),
with periodicity $1/T$ \cite{kapusta}.
This leads to a discrete spectrum for the zero component of the momentum
$q_0$
\be
q_0 \rightarrow 2 \pi m T,~~~~~~~~~m=0,\pm1,\pm2,...
\label{fourone} \ee
As a consequence the integration over $q_0$ is replaced by
a summation over the
discrete spectrum
\be
\int \frac{d^d q}{(2 \pi)^d} \rightarrow
T \sum_m \int \frac{d^{d-1}\vec{q}}{(2 \pi)^{d-1}}.
\label{fourtwo} \ee

With the above remarks in mind we can easily generalize our master equation
(\ref{twonine}) in order to take into account the temperature effects.
For the temperature dependent effective
average potential $U_k(\rhoa,\rhob,T)$
we obtain
\be
\frac{\partial}{\partial t} U_k(\rhoa,\rhob,T) =
\frac{1}{2} (2 \pi)^{-(d-1)} T \sum_m \int d^{d-1} \vec{q}~~
\left( \frac{1}{P + M^2_1}
+\frac{1}{P + M^2_2} \right)
\frac{\partial}{\partial t} R_k,
\label{fourthree} \ee
with the implicit replacement
\be
q^2 \rightarrow \vec{q}^2 + 4 \pi^2 m^2 T^2
\label{fourfour} \ee
in eqs. (\ref{twotwo}),(\ref{twothree}) and (\ref{twothirteen})
for $R_k$ and $P$.
Again, the usual temperature dependent effective potential
\cite{doljack}-\cite{linde}
is obtained from $U_k(\rhoa,\rhob,T)$ in the limit $k \rightarrow 0$.
As before, we can parametrize $U_k(\rhoa,\rhob,T)$ in terms of its minimum and
its derivatives at the minimum. The evolution equations are given by
(\ref{threetwo})-(\ref{threefour}),
(\ref{threeseven})-(\ref{threenine}),
(\ref{threethirteen})-(\ref{threefifteen}),
with the obvious generalizations
\beq
\rhz(k)  \rightarrow & \rhz(k,T) \nonumber \\
\lb(k)    \rightarrow & \lb(k,T)   \nonumber \\
x(k)    \rightarrow & x(k,T)   \nonumber \\
\mb(k)    \rightarrow & \mb(k,T).
\label{fourfive} \eeq

The $L^d_n$ integrals for non-vanishing temperature read
\be
L^d_n(w,T) =
- 2 n k^{2n-d} \pi^{-\frac{d}{2}+1} \Gamma \left( \frac{d}{2} \right)
 T \sum_m
\int d^{d-1} \vec{q}~~ \frac{\partial P}{ \partial t} (P + w)^{-(n+1)},
\label{foursix} \ee
where the implicit replacement
(\ref{fourfour}) is assumed in $P$.
Their basic properties can be established
analytically.
For $T \ll k$ the summation over discrete
values of $m$ in the expression (\ref{foursix}) is equal to the
integration over a continuous range of $q_0$ up to exponentially small
corrections.
Therefore
\be
L^d_n(w,T) = L^d_n(w)~~~~~~~~~{\rm for}~~T \ll k.
\label{fourseven} \ee
In the opposite limit $T \gg k$ the summation over $m$ is
dominated by the $m=0$ contribution.
Terms with non-zero values of $m$ are suppressed
by $  \sim \exp \left( -( mT/k)^{2 } \right).$
The leading contribution gives the
the simple expression
\be
L^d_n(w,T) = \frac{v_{d-1}}{v_d} \frac{T}{k} L^{d-1}_n(w)~~~~~~{\rm
for}~~T \gg k,
\label{foureight} \ee
with $v_d$ defined in (\ref{threevd}).
The two regions of $T/k$ in which $L^d_n(w,T)$ is given by the
equations (\ref{fourseven}), (\ref{foureight})
are connected by a small interval,
in which
the exponential corrections result in a more
complicated dependence on
$w$ and $T$.
\par
The above conclusions are verified by a numerical calculation of
$L^4_1(w,T)$.
In fig. 3 we plot
\footnote{Comparison with results presented in
ref. \cite{transition} shows that the form of these functions
depends on the details of the infrared regulating function
$R_k(q)$. However, the physical results,
which are obtained when the cutoff is removed, are independent of
the shape of the cutoff. This will be apparent in the
next sections and is a verification of the scheme independence
of our conclusions.
}
$L^4_1(w,T)/L^4_1(w)$ as a function of $T/k$,
for various values of $w/k^2$.
We distinguish three regions: \\
a)~$ T/k \leq \theta_1$: This is the
\underline{low temperature region} where $L^4_{1,2}(w,T)$
are very well approximated by their
zero temperature value.
We take $\theta_1=0.15$ and use $L_{1,2}^4(w,0)$ in the evolution equations
for $k \geq T/ \theta_1$.
\\
b)~$\theta_1 <T/k< \theta_2$:
In the \underline{threshold region}
we perform a numerical
fit of the curve corresponding to
$w=0$ which we use for all values of $w$. This is
a good
approximation
since the relevant $w/k^2$ turns out to be small in this region (see next
sections). \\
c)~$ T/k \geq \theta_2$:
We take $\theta_2= 0.4$. For the
\underline{high temperature region} we use for
the numerical solution of the evolution equations
\be
L^4_{1,2}(w,T)= 4 \frac{T}{k} L^3_{1,2}(w).
\label{fournine} \ee
The three dimensional character of the effective theory for modes
with $q^2 \ll T^2$ manifests itself in the appearance of the three dimensional
momentum integrals. It acquires here a precise quantitative meaning.

We have now developed the necessary formalism
for the study of the four-dimensional
zero and non-zero temperature theory.
In the following two sections
we study the evolution of the
running parameters of the theory, which
leads to the determination of the renormalized theory at
zero and non-zero temperature.

\setcounter{equation}{0}
\renewcommand{\theequation}{{\bf 6.}\arabic{equation}}

\section*{6.
The running in the low temperature and threshold
regions}

In section 4 we derived the zero temperature evolution equations
for the parameters (masses,
vacuum expectation values and couplings)
of the truncated theory as a function of the scale $k$
in the various regimes (S, M, AX). In section 5 we
generalized the formalism in order to take into account
non-zero temperature effects.
The evolution equations can be solved for a given
set of initial conditions, specified as the values of the
running parameters at a scale equal to the
ultraviolet cutoff of the theory ($k = \Lambda$).
As we pointed out in section 2, at this scale the effective
average action is equal to the classical action. Therefore,
the initial values for the parameters correspond to
their classical (or bare) values.
Also, the discussion in section 5 has shown that
in the low temperature region ($k \ge T/ \theta_1$)
there is no difference between the zero and non-zero temperature
theory. As a result, we can define the theory in
terms of the classical values of its parameters at
$k = \Lx \gg T$, independently of the temperature.
The integration of the evolution equations gives the running
couplings at lower scales. No temperature effects are observed
in the evolution inside the low temperature region
($k \ge T/ \theta_1$). We can, therefore, use the values of the
running couplings at $k = T/ \theta_1$ for the definition of
the theory, since they are in one-to-one correspondence
with the classical couplings, independently of the temperature.
This turns out to be the most convenient choice and we
shall use it for the rest of the paper.
The temperature starts to become
important when the evolution enters the threshold region
($T/ \theta_1 > k > T/ \theta_2$).
In the high temperature region
($k \le  T/ \theta_2$) the evolution is effectively three-dimensional,
as we discussed in section 5.
Finally in the limit $k \rightarrow 0$ the effective average action
becomes the effective action, and the integration of the evolution
equations gives the renormalized values for the couplings at
various temperatures.
All the information on the various phases of the theory is
contained in these renormalized couplings and their
temperature dependence.

We have seen in the introduction and in section 4 that
the AX regime ($x>0$) and the M regime
($x<0$) can be mapped onto each other
through a simple
redefinition of the fields (see eqs. (\ref{threeeleven}),
(\ref{threetwelve})).
For this reason, the physical behaviour in the two regimes
is the same. For example, the
Cubic and Ising fixed points generate the same
universal behaviour (characteristic of a $Z_2$-symmetric scalar theory).
For this reason, we shall concentrate on the region
$x>0$ only. All the results can be easily extended to
the region $x<0$, through the transformations
of eqs. (\ref{threeeleven}), (\ref{threetwelve}).

Since we are interested in symmetry restoration at non-zero
temperature,
we first consider the theory in the spontaneously broken regime.
The evolution equations in the AX regime
(which is the relevant one for $x>0$)
in four dimensions and
non-zero temperature can be easily derived from eqs.
(\ref{threethirteen})-(\ref{threefifteen})
and read
\beq
\frac{d \rhz}{d t} =~& - v_4 k^2
\bigl\{ 3 L_1^4(2\lb \rhz) t_1(2 \lb \rhz, T)
+ (1+x)  L_1^4(x \lb \rhz)  t_1(x \lb \rhz, T)
\bigr\}
\label{sixone}
\\
\frac{d \lb}{d t} =~& -v_4
\lb^2
\bigl\{ 9 L_2^4(2 \lb \rhz)  t_2(2 \lb \rhz, T)
+ (1+x)^2  L_2^4(x \lb \rhz)  t_2(x \lb \rhz, T)
\bigr\}
\label{sixtwo}
\\
\frac{d x}{d t} =~& v_4
\frac{6}{\rhz}  \frac{x+\frac{x^2}{4}}{1-\frac{x}{2}}
\bigl\{ L_1^4(2\lb \rhz)  t_1(2 \lb \rhz, T)
- L_1^4(x \lb \rhz)  t_1(x \lb \rhz, T)
\bigr\}
\nonumber \\
&+ v_4 x \lb
\bigl\{
9 L_2^4(2 \lb \rhz)  t_2(2 \lb \rhz, T)
+ (1+x)^2
L_2^4(x \lb \rhz)  t_2(x \lb \rhz, T)
\bigr\}.
\label{sixthree}
\eeq
where $v_4 =1/32 \pi^2$. We have not indicated explicitly
the $k$ and $T$
dependence of the running parameters
$\rhz(k,T)$, $\lb(k,T)$, $x(k,T)$.
They are defined
at zero temperature according
to eqs. (\ref{threeten}), and generalized for non-zero temperature
according to eqs. (\ref{fourfive}).
The functions $t_{1,2}$ are defined as
\be
t_n(w,T) =  \frac{L^4_n(w,T)}{L^4_n(w)},
\label{sixfour} \ee
with $t_1(w,T)$ plotted in fig. 3.
\par
At zero temperature one has $t_n(w,0) =1$
and the evolution equations have
only one infrared attractive fixed point, the Gaussian one.
In the limit of small
$\lb$ we shall neglect the slow logarithmic running of $\lb$,
which is eventually stopped by the mass terms in the
threshold functions $L^4_{1,2}$.
Similarly the running of $x$ can also be neglected
since it is suppressed by $\lb/32 \pi^2$. (For small $\lb$ the
difference of the two $L^4_1$ functions in the first line of
eq. (\ref{sixthree}) gives a contribution $\propto \lb L^4_2(0)$.)
For large $\lb$ the evolution equations can be integrated numerically
and the small resulting corrections can be reliably computed.
This has been done in ref. \cite{transition} for the
$O(N)$-symmetric scalar theory. In this paper we
concentrate on small couplings for which
analytical expressions can be obtained.
Eq. (\ref{sixone}) can be integrated easily for small $\lb$
and we obtain ($L^4_1(0)= -2$)
\be
\rhz(k,0) = \rhz \left( \frac{T}{\theta_1} \right)
+  \frac{1}{32 \pi^2}
(x+4) \left[ k^2 - \left(\frac{T}{\theta_1} \right)^2 \right],
\label{sixfive} \ee
where we used the point $k = T/\theta_1$ instead of
$k=\Lx$ to start the evolution, as we have
explained in the first paragraph of this section.
We define the renormalized couplings of the theory in
the limit $k \rightarrow 0$ as
\footnote{In the case that Goldstone modes
are present (as for $x=0$) the couplings are defined at
some appropriate non-zero $k$.
The same applies for non-zero temperature. This does not affect
our results for small $\lb$. For a detailed discussion
see ref. \cite{transition}.}
\beq
\rhz =& \rhz(0,0)
\nonumber \\
\lr =& \lb(0,0)
\nonumber \\
\xr =& x(0,0).
\label{sixsix} \eeq
and conclude that
\be
\rhz \left( \frac{T}{\theta_1} \right)
= \rhz
+  \frac{1}{32 \pi^2}
(x_R+4) \left( \frac{T}{\theta_1} \right)^2.
\label{sixseven} \ee
\par
At non-zero temperature, the evolution in the
low temperature region ($k \ge T/ \theta_1$) is
identical to the zero temperature case.
In the threshold region
($T/ \theta_1 > k > T/ \theta_2$), the form of $t_{1,2}(w,T)$
is not given by a simple analytical expression.
For small $\lb$ we neglect the running of
$\lb, x$ in this region and find
\beq
\rhz\left( \frac{T}{\theta_2}, T \right)
=& \rhz \left( \frac{T}{\theta_1} \right)
-  \frac{1}{16 \pi^2}
(x+4) T^2 I
\nonumber \\
=& \rhz +  \frac{1}{16 \pi^2} (x_R+4) T^2
\left( \frac{1}{2 \theta_1^2} - I \right)
\nonumber \\
\lb \left( \frac{T}{\theta_2}, T \right)
=& \lb \left( \frac{T}{\theta_1} \right) = \lr
\nonumber \\
x \left( \frac{T}{\theta_2},T \right)
=& x \left( \frac{T}{\theta_1} \right) = \xr,
\label{sixeight} \eeq
where
\be
I = \int_{1/\theta_2}^{1/\theta_1}
dy y ~t_1 \left( 0,\frac{1}{y} \right)
\label{sixnine} \ee
and we have made use of the fact that
$t^4_1(w,T)$ depends on $T$ only through the
combination $T/k$. The integral $I$ can be evaluated
numerically. For $\theta_1=0.15$, $\theta_2=0.4$ we
find $I=19.97$.
Eqs. (\ref{sixeight})
set the initial values for
the evolution in the high temperature region.

\setcounter{equation}{0}
\renewcommand{\theequation}{{\bf 7.}\arabic{equation}}

\section*{7.
The running in the high temperature
region}

In the high temperature region ($k \le T/\theta_2$) the functions
$L^4_{1,2}(w,T)$ are given by the simplified expression
(\ref{fournine}).
We can rewrite eqs. (\ref{sixone})-(\ref{sixthree})
in terms of effective three-dimensional couplings
\beq
\rhz'(k,T) =& \frac{\rhz(k,T)}{T}
\nonumber \\
\lb'(k,T) =& \lb(k,T) T
\nonumber \\
\gb'(k,T) =& \gb(k,T) T
\nonumber \\
x(k,T) =& \frac{\gb'(k,T)}{\lb'(k,T)} -1.
\label{sixeleven} \eeq
The flow equations then read
\beq
\frac{d \rhz'}{d t} =~& - v_3 k
\bigl\{ 3 L_1^3(2\lb' \rhz') + (1+x)  L_1^3(x \lb' \rhz') \bigr\}
\label{sixtwelve}
\\
\frac{d \lb'}{d t} =~& -v_3 k^{-1}
[\lb']^2
\bigl\{ 9 L_2^3(2\lb' \rhz') + (1+x)^2  L_2^3(x \lb' \rhz') \bigr\}
\label{sixthirteen}
\\
\frac{d x}{d t} =~& v_3 k^{-1}
\frac{6}{\rhz'}  \frac{x+\frac{x^2}{4}}{1-\frac{x}{2}}
\bigl\{ L_1^3(2\lb' \rhz') - L_1^3(x \lb' \rhz') \bigr\}
\nonumber \\
&+ v_3 k^{-1} x \lb'
\bigl\{
9 L_2^3(2 \lb' \rhz')
+ (1+x)^2
L_2^3(x \lb' \rhz') \bigr\},
\label{sixfourteen}
\eeq
with $v_3 = 1/8 \pi^2$.
Comparison with eqs. (\ref{threethirteen})-(\ref{threefifteen})
shows that the above equations are exactly the ones of the
three-dimensional theory at zero temperature.
In order to make their fixed point structure
more transparent we
define the dimensionless parameters
\beq
\kx(k,T) =& \frac{\rhz'(k,T)}{k} = \frac{\rhz(k,T)}{k T}
\nonumber \\
\lx(k,T) =& \frac{\lb'(k,T)}{k} = \frac{\lb(k,T) T}{k}
\nonumber \\
g(k,T) =& \frac{\gb'(k,T)}{k} = \frac{\gb(k,T) T}{k}.
\label{sixfifteen} \eeq
In terms of these we obtain the scale invariant form of the
evolution equations
\beq
\frac{d \kx}{d t} =~& -\kx - v_3
\bigl\{ 3 L_1^3(2\lx \kx) + (1+x)  L_1^3(x \lx \kx) \bigr\}
\label{sixsixteen}
\\
\frac{d \lx}{d t} =~& -\lx -v_3
\lx^2
\bigl\{ 9 L_2^3(2\lx \kx) + (1+x)^2  L_2^3(x \lx \kx) \bigr\}
\label{sixseventeen}
\\
\frac{d x}{d t} =~& v_3
\frac{6}{\kx}  \frac{x+\frac{x^2}{4}}{1-\frac{x}{2}}
\bigl\{ L_1^3(2\lx \kx) - L_1^3(x \lx \kx) \bigr\}
\nonumber \\
&+ v_3 x \lx
\bigl\{
9 L_2^3(2 \lx \kx)
+ (1+x)^2
L_2^3(x \lx \kx) \bigr\}.
\label{sixeighteen}
\eeq
No explicit dependence on the scale $k$ appears on the r.h.s.
\par
The first of the above equations defines a critical surface
$\kx_{cr}=\kx_{cr}(\lx,x)$.
It consists of the points
$\kx_{cr}$ for which the solution of
eqs. (\ref{sixsixteen})-(\ref{sixeighteen}) approaches, for large
negative $t$, a scaling solution with $\kx$, $\lx$ and $x$
independent of $t$.
(For a weakly first order transition the scaling holds only
approximately.)
Every point on the critical surface is unstable
in the $\kx$ direction.
Trajectories
which start at $\kx> \kx_{cr}$ continue towards the region
of large $\kx$, in such a way that $\rhz(k,T)=\kx(k,T) T k$
reaches asymptotically a constant value for $k \rightarrow 0$.
As a result the renormalized theory settles down in the
phase with spontaneous symmetry breaking.
If the evolution starts at
$\kx < \kx_{cr}$, the flows cross the surface
$\kx = 0$ at some finite $k_s$. From this point on
the system is in the symmetric regime. In order to
continue the evolution, we define appropriate parameters
according to eqs. (\ref{lala}).
The evolution equations read
\beq
\frac{d {\bar m}^2}{d t} &= v_3 k  (4+x)  \lb' L_1^3({\bar m}^2)
\label{extone} \\
\frac{d \lb'}{d t} &= - v_3 k^{-1} (10+2x+x^2)
[\lb']^2 L_2^3({\bar m}^2)
\label{exttwo} \\
\frac{d x}{d t} &= v_3 k^{-1} (x+1)x(x-2)
\lb'  L_2^3({\bar m}^2),
\label{extthree}
\eeq
with
$\lb'$ related to $\lb$ by eqs. (\ref{sixeleven}).
Comparison  with eqs. (\ref{threetwo}) - (\ref{threefour})
shows that the above equations are the ones for the
three-dimensional theory in the symmetric regime.
We define the dimensionless couplings
\beq
m^2(k,T) =~& \frac{\mb(k,T)}{k^2}
\nonumber \\
\lx(k,T) =~& \frac{\lb'(k,T)}{k} = \frac{\lb(k,T) T}{k}
\nonumber \\
g(k,T) =~& \frac{\gb'(k,T)}{k} = \frac{\gb(k,T) T}{k}.
\label{extfive} \eeq
In terms of these quantities the evolution equations in the
symmetric regime read
\beq
\frac{d m^2}{d t} &= -2 m^2 + v_3   (4+x)  \lx L_1^3(m^2)
\label{extsix} \\
\frac{d \lx}{d t} &= -\lx - v_3  (10+2x+x^2)
\lx^2 L_2^3(m^2)
\label{extseven} \\
\frac{d x}{d t} &= v_3 (x+1)x(x-2)
\lx  L_2^3(m^2).
\label{exteight}
\eeq
We start the evolution in this regime at $k = k_s$ with
$m^2(k_s,T)=0$ and $\lx(k_s,T)$, $x(k_s,T)$
taking their values at the end of the running in the
spontaneously broken regime.
For $k \rightarrow 0$ the evolution is stopped by the
mass terms in the threshold functions $L^3_{1,2}$
and the theory settles down in the symmetric phase.
Obviously the critical temperature $\tcr$ is related to
$\kx_{cr}$ (for given $\lx(\tcr/\theta_2)$, $x(\tcr/\theta_2)$)
by
\be
\kx \left( \frac{\tcr}{\theta_2} \right) = \kx_{cr}.
\label{chrst} \ee
\par
On the critical surface there are two
fixed points
with at least one attractive direction for the flow towards the
infrared ($k \rightarrow 0)$: \\
a) the Cubic fixed point located at
\be
\kx_C=5.674 \times 10^{-2}~~~~\lx_C=8.747~~~~x_C=2,
\label{sixnineteen} \ee
b) and the Heisenberg fixed point located at
\be
\kx_H=4.486 \times 10^{-2}~~~~\lx_H=15.265~~~~x_H=0.
\label{sixtwenty} \ee
Both fixed points
are attractive in the $\lx$ direction,
but the Cubic fixed point is unstable in the
$x$ direction
while the Heisenberg one is stable.
For fixed $x$ there is also the infrared unstable
Gaussian fixed point. It is located at
\be
\kx_G=-v_3(x+4)L^3_1(0)=\frac{1}{8 \pi^{3/2}}(x+4)~~~~\lx_G=0.
\label{sixtwentyone} \ee
These fixed points are the same as the ones observed in section
3. The only difference lies in the parametrization. In section
3 we used an expansion around the origin of the potential
and the evolution equations in the symmetric regime. For this
reason the fixed points appeared for negative values of
the mass parameter (the curvature at the origin), indicating
a minimum away from the origin. In this section we rely
on a parametrization around the minimum of the potential, which
results in increased quantitative accuracy for the truncation.
\par
The flows on the critical surface are qualitatively similar to those
in fig. 1 for $x>0$. There are two disconnected regions: \\
a)$2 > x >0$: The trajectories flow away from the Cubic fixed point
and towards the Heisenberg fixed point. This region corresponds to
a second order phase transition.\\
b)$x >2$: The trajectories flow away from the Cubic fixed point
and into a region of small $\lx$ and large $x$. Similarly to our
discussion in section 3, we expect
$\lx(k,T)$ to become negative at
some finite $k$. This indicates that the minimum of the potential
becomes unstable (it turns into a maximum).
Our crude truncation is not sufficient for the investigation
of this situation, since the higher derivatives of the potential
are important. A detailed study is presented in ref.
\cite{stefan} for the four-dimensional theory at zero
temperature. The dimensionality of the theory is not
crucial for the qualitative behaviour
in this region of parameter space. The first term
in the r.h.s of eq. (\ref{sixseventeen})
(which is present only for the effectively three-dimensional theory)
is not important
for large $x$. The second term, which drives the
dynamics, is the same as for the four-dimensional theory
(with the replacement of $v_3 L^3_1$ by  $v_4 L^4_1$
generating only quantitative corrections.)
In ref. \cite{stefan} higher derivatives of the potential
are taken into account. Also a parametrization is used which
simultaneously
follows the evolution of the potential at its minimum at non-zero
$\rhz(k,T)$ and at the origin.
This permits the study of the global properties of the potential.
It is found that during the evolution in the region
$x>2$
a second minimum appears at the origin which subsequently
becomes the absolute minimum of the potential.
This results in a discontinuity in the order parameter and a
first order phase transition. At some point in the evolution,
$\lx$ turns negative
and the minimum at non-zero $\rhz$ disappears. From this point
on the deeper minimum at zero is the only minimum.
During the whole evolution the higher derivatives stay positive
guaranteeing that the potential remains bounded.
We cannot reproduce the above picture within the
crude truncation of a quartic polynomial for the
potential. Instead we shall
give in section 11
an approximate solution of the evolution equation for the
potential in this region, which will demonstrate the
existence of the first order transition.

\setcounter{equation}{0}
\renewcommand{\theequation}{{\bf 8.}\arabic{equation}}

\section*{8.
The critical temperature}

A quantity which can be easily calculated from the discussion in the
last two sections is the critical temperature for the
phase transitions. From eqs. (\ref{sixeight}) and (\ref{sixfifteen})
we obtain
\beq
\kx \left( \frac{T}{\theta_2}, T \right)
=& \theta_2 \frac{\rhz}{T^2} +  \frac{\theta_2}{16 \pi^2}
\left( \frac{1}{2 \theta_1^2} - I \right) (x_R+4)
\nonumber \\
\lx \left( \frac{T}{\theta_2}, T \right)
=& \theta_2 \lr
\nonumber \\
x \left( \frac{T}{\theta_2}, T \right)
=&  \xr.
\label{sixtwentytwo} \eeq
For small $\lx$ the critical surface, which separates the symmetric
phase from the phase with spontaneous symmetry breaking,
goes through the Gaussian fixed point given by eq.
(\ref{sixtwentyone}).
The critical temperature can be computed as the temperature
for which
$\kx \left( \frac{T}{\theta_2}, T \right)$ coincides with the
Gaussian fixed point $\kx_G$.
This gives
\beq
\frac{\tcr^2}{\rhz} =& \frac{C}{x_R+4}
\nonumber \\
C^{-1} =& \frac{1}{8 \pi^2} \left[
\frac{\sqrt{\pi}}{\theta_2}
- \frac{1}{2} \left( \frac{1}{2 \theta_1^2} - I \right)
\right],
\label{sixtwentythree} \eeq
independently of $\lr$.
Substitution of the values $\theta_1=0.15$, $\theta_2=0.4$,
$I=19.97$, which we computed in part I of this section,
gives
\be
C=23.89.
\label{sixtwentyfour} \ee
We should point out that the above value for the critical
temperature is not strictly valid for $x_R>2$. In this region
the first order phase transition occurs for
$\kx \left( \frac{T}{\theta_2}, T \right)$
slightly above the critical surface.
As a result the transition takes place at a temperature
slightly lower than the one given by eq. (\ref{sixtwentythree}).
In the language of the effective three-dimensional theory,
the distance from the phase transition
can be parametrized,
for small $\lr$,
in terms of
$\dkcr = \kx \left( \frac{T}{\theta_2}, T \right) - \kx_G$.
We establish the connection between this quantity and the
temperature as
\be
\dkcr = \kx \left( \frac{T}{\theta_2}, T \right) - \kx_G
= \theta_2 \rhz \left( \frac{1}{T^2} - \frac{1}{\tcr^2} \right).
\label{sixtwentyfive} \ee
\par
The critical temperature can also be calculated
in high temperature perturbation theory through
the perturbative expansion of the effective potential
\cite{colwein} and its generalization to
non-zero temperature \cite{doljack}-\cite{linde}.
The calculation is straightforward and we do not present
the details here. When the leading term in the high temperature
expansion of the one-loop contribution
to the effective potential is retained,
the critical temperature is found to be
\be
\frac{\tcr^2}{\rhz} = \frac{24}{x_R+4}.
\label{threetwentysix} \ee
This value is in excellent agreement with our result.
The slight discrepancy is due to
small deviations of the form of $L^4_1(w,T)$ that we have used
from the exact expression, and could probably be removed by
using lower $\theta_1$ and larger $\theta_2$.
It is well known that the perturbative expansion
of the effective potential breaks down near the critical
temperature for a second or weakly first order phase transition,
due to infrared divergences
\cite{transition}.
The surprising accuracy of the perturbative estimate for
the critical temperature is due to the fact that the
infrared divergences appear at temperatures
$|T-\tcr|/\tcr = {\cal{O}} (\lr)$.
For sufficiently small $\lr$ the location of the transition can
be accurately computed. However, naive perturbative predictions
for the details of the transition
(even its order) can be misleading. This is the case for our model,
for which the  next terms in the naive
high temperature expansion of the
perturbative result fail to reproduce even the correct
qualitative picture
\footnote{Perturbation theory for gap equations
\cite{buch} may lead to more reliable results concerning the order
of the transition but will fail for critical exponents unless
the effectively three-dimensional running of the quartic scalar
coupling
is properly included}
(apart from a small region in parameter space
which will be discussed at the end of the next section).
We emphasize that our approach may also be used for large values
of $\lr$ not so easily accessible to perturbation theory. In this
case $\kx_G$ should be replaced by the relevant
exact point on the critical
surface $\kx_{cr}$.

\setcounter{equation}{0}
\renewcommand{\theequation}{{\bf 9.}\arabic{equation}}

\section*{9.
Second order phase transition and the critical behaviour}

Let us briefly summarize the main results of the
previous sections. We have considered
theories which at zero temperature are
in the phase with
spontaneous symmetry breaking corresponding
to the AX regime.
They are defined in terms of the classical
(bare) parameters at the ultraviolet cutoff
$\Lambda \gg T$.
The renormalized parameters are obtained
by solving the evolution equations from $k=\Lx$ to
$k=0$.
In the low temperature region
($\Lambda \geq k \geq T/\theta_1$)
there is no difference between the zero and
non-zero temperature case
for the evolution of
the parameters of the theory.
The first temperature effects are observed in the
threshold region ($T/\theta_1 > k > T/\theta_2$).
For small $\lx_R$
the values of the running parameters
at the beginning of the evolution in the
high temperature region ($k = T/\theta_2$)
can be expressed
in terms of the renormalized parameters
of the zero temperature theory.
The relation is given by eqs. (\ref{sixeight}).
In the high temperature region
($k \leq T/\theta_2$) the character of the
evolution is effectively three-dimensional
and is governed by the fixed points of the
three-dimensional theory.
These are most transparent in terms of the
dimensionless parameters defined
in eqs. (\ref{sixfifteen}). The evolution
equations are given by eqs.
(\ref{sixsixteen})-(\ref{sixeighteen}).
These equations define a critical surface
$\kx_{cr} = \kx_{cr}(\lx,x)$, which is
unstable in the $\kx$ direction and separates the
phase with spontaneous symmetry breaking
from the symmetric one.
The system
ends up in either phase depending on the
values of the running parameters at
$k = T/\theta_2$.
For small $\lx$
the critical surface
goes through the Gaussian fixed point given
by eq. (\ref{sixtwentyone}).
As a result, for small $\lx_R$ the crucial quantity is
the distance from the critical surface
$\dkcr = \kx \left( \frac{T}{\theta_2}, T \right) - \kx_G$.
For $\dkcr > 0$ the theory ends up in the
phase with spontaneous symmetry breaking.
For $\dkcr < 0$ it ends up in the symmetric one.
There is a direct connection between $\dkcr$ and
the distance from the critical temperature, which
is expressed in eq. (\ref{sixtwentyfive}).

In this section we discuss the region in parameter space
$0<x<2$ where the phase transition is second order.
Sufficiently near to the critical temperature the evolution
is governed by the Heisenberg fixed point.
In fig. 4 we plot the numerical solution of eqs.
(\ref{sixsixteen})-(\ref{sixeighteen}) for the evolution
in the high temperature region, for a theory with
zero temperature renormalized parameters
$\lr=0.01$, $\xr=1$ and critical temperature
$\tcr^2/\rhz=4.78$.
We display two trajectories, which
start slightly
above and below the critical surface (and therefore
correspond to temperatures slightly below and above
the critical one).
We observe that the system flows towards the
Heisenberg fixed point which is attractive in both the
$\lx$ and $x$ directions. It stays around this fixed point
for several orders of magnitude in $t$ and then
deviates towards either the phase with spontaneous symmetry
breaking or the symmetric one. During the ``time'' $t=\ln(k/\Lx)$
that the system stays close the fixed point it loses
memory of the initial conditions of the evolution.
Its dynamics is fixed solely by the fixed point, which
has a purely three-dimensional character (as we demonstrated
in section 7). As a result
we expect that the behaviour of the theory near the critical
temperature is independent of the
details of the zero temperature theory. It must display
universal critical behaviour characteristic of systems with
Heisenberg fixed point.
As long as $\kx(k,T)$ stays almost constant around its fixed point
value $\kx_H$
and $k\rightarrow 0$ as $t \rightarrow -\infty$,
the minimum of the effective average potential
evolves towards zero according to
\be
\rhz(k,T)=\kx_H k T.
\label{nineone} \ee
If the temperature is equal to the critical one, the system
never leaves the fixed point and $\rhz(0,\tcr) = 0$.
If the temperature is slightly below $\tcr$, $\kx(k,T)$
eventually runs away from the fixed point and diverges,
so that
$\rhz(k,T)=\kx(k,T) k T$ reaches a constant non-zero value
as $k\rightarrow 0$.
This value corresponds to the renormalized minimum of the
effective potential at non-zero temperature
and we denote it by
\be
\rhz(T) = \rhz(0,T).
\label{ninetwo} \ee
For a temperature slightly above $\tcr$,
$\kx(k,T)$ (and therefore $\rhz(k,T)$)
runs to zero at a finite $k_s$. From this
point on the system is in the symmetric regime and the
appropriate evolution equations are given by
eqs. (\ref{extsix})-(\ref{exteight}).
We start the evolution in this regime at $k = k_s$ with
$m^2(k_s,T)=0$ and $\lx(k_s,T)$, $x(k_s,T)$
taking their values at the end of the running in the
spontaneously broken regime.
For $k \rightarrow 0$ the evolution is stopped by the
mass terms in the threshold functions $L^3_{1,2}$
and the theory settles down in the symmetric phase.
We define the renormalized mass in the symmetric phase
as
\be
\mr(T) = \mb(0,T).
\label{ninethree} \ee
We also define the renormalized couplings in both
phases as
\beq
\lr (T) = \lb(0,T) \nonumber \\
\xr (T) = x(0,T).
\label{ninefour} \eeq
It is important to point out that, while the system is staying
close to the fixed point, the coupling $\lb(k,T)$
evolves towards zero according to
\be
\lb(k,T)=\lx_H \frac{k}{T}.
\label{ninefive} \ee
As a result $\lr(T)$ goes to zero as the critical
temperature is approached.
Its strong renormalization near $\tcr$ provides the
resolution of the problem of infrared divergences.
The ratio $\lr(T) T/m_R(T)$ does not diverge
near the critical temperature,
in contrast to $\lr T/m_R(T)$. (Here
$\lr$ is the renormalized coupling
of the zero-temperature theory, which is approximately
equal to the bare one for small couplings.)
We shall not elaborate on this point, but we refer the reader
to ref. \cite{transition} for an extensive discussion.
We also mention that as the temperature deviates
from the critical one the system spends less ``time''
$t= \ln(k/\Lx)$
near the critical point. Its flow deviates from the those
depicted in fig. 4 at earlier stages.
As a result the universal behaviour ceases to dominate.

The behaviour of the renormalized theory at various
temperatures is shown in fig. 5
for zero temperature parameters
$\lr=0.01$, $\xr=1$ and critical temperature
$\tcr^2/\rhz=4.78$.
We observe that $\rhz(T)$ moves continuously to zero,
indicating a second order phase transition.
The mass $\mr (T)$ is zero at $\tcr$ and positive for
larger temperatures.
The quartic coupling
$\lr(T)$ stays close to its zero temperature value
for most temperatures, but is strongly renormalized
towards zero near $\tcr$.
The ratio of couplings
$\xr (T)$ again takes its zero temperature value,
unless the temperature is sufficienlty close to
$\tcr$ for the flow to reach the Heisenberg fixed point.
The universal behaviour near $\tcr$ can be parametrized by
critical exponents, which we define
similarly to ref. \cite{transition} as
\beq
\rhz(T) \propto& (\tcr^2 - T^2)^{2 \beta} \nonumber \\
\mr(T) \propto& (T^2-\tcr^2)^{2 \nu} \nonumber \\
\lr(T) \propto& (T^2-\tcr^2)^{\zeta} \nonumber \\
\xr(T) \propto& (T^2-\tcr^2)^{\mu}.
\label{ninesix} \eeq
More precisely,
$2 \beta$ is given by the derivative of $\ln \rhz(T)$
with respect to $\ln(\tcr^2-T^2)$, and similarly for the
other parameters.
The definiton of $\zeta$ and $\mu$
applies only to the symmetric phase.
The exponent $\beta(T)$ is plotted in fig. 6 along with
$\xr(T)$ for temperatures approaching $\tcr$, for a
theory with $\lx_R=0.2$, $x_R=1$.
It is apparent from the temperature dependence
of $\xr$ that near $\tcr$ the Heisenberg
fixed point becomes important. During its evolution
the system stays long
enough on the critical surface for this fixed point to
generate universal critical behaviour.
The exponent $\beta(T)$ approaches a temperature independent
value which is independent of $\lr$ and $x_R$
(as long as $x_R < 2$) and characteristic of systems with
Heisenberg critical behaviour. This value is
\be
\beta_H = 0.32
\label{nineseven} \ee
in agreement with ref. \cite{transition}.
Two other exponents are fixed by the scaling laws and
the finite value of the ratio $\lr(T) T/m_R(T)$. They
are
$\nu_H = \zeta_H = 2 \beta_H$.
The above values for the exponents are in rough agreement with
known values from three-dimensional field theory
\cite{amit,zinn}. The agreement
improves dramatically when less restrictive truncations are
used for the study of
the evolution equation for the potential, and wave function
renormalization effects are taken into account \cite{indices}.
We have performed this more accurate calculation and
obtained results which agree with the known values at the
4-5 \% level. This work will be described in ref. \cite{peter}.
Finally, the exponent $\mu$ also approaches asymptotically a
constant value (c.f. eq. (\ref{sixeighteen}))
\beq
\mu_H =& \frac{\nu_H}{8 \pi^2}
\biggl\{
\frac{6}{\kx_H} \left[ L_1^3(2\lx_H \kx_H) - L_1^3(0) \right]
+ \lx_H \left[ 9 L_2^3(2 \lx_H \kx_H) + L_2^3(0) \right]
\biggr\} \nonumber \\
=& 0.64.
\label{expmu} \eeq

\setcounter{equation}{0}
\renewcommand{\theequation}{{\bf 10.}\arabic{equation}}

\section*{10. Tricritical point and crossover}

In the above discussion the Heisenberg fixed point
was the only one which played any role. This was
expected since the Cubic fixed point
is repulsive in the $x$ direction. Any flow which starts
sufficiently far from it is further repelled and the system
never feels its effect.
However, it is possible that the values
of the running parameters at the beginning of the
evolution in the high temperature region are
within the region of influence of the Cubic
fixed point. An example is given in fig. 7, for a
theory with
$\lr=0.01$, $\xr$ slightly smaller than 2, and
$\tcr^2/\rhz=3.98$. As we have discussed in
the introduction and section 3, flows that start
on the surface $x=2$ in parameter space never
move out of it. The flows depicted in fig. 7
start with
$x \left( \frac{T}{\theta_2}, T \right) = 2 -\dx$
and $\dx \ll 1$.
For this reason, their deviation
from the surface $x=2$ is very slow.
We display two trajectories which start a small
distance $\dkcr$ above and below the critical surface
(and therefore correspond to temperatures
slightly below and above the critical one, according
to eq. (\ref{sixtwentyfive})).
For $|\dkcr| \ll \dx \ll 1$ the flows stay on
the critical surface and close to $x=2$ for
a large initial part of the evolution. During
this ``time'' they approach the Cubic fixed point
and stay near it. Finally $x(k,T)$ starts
growing and the system moves
away from the repulsive (in the $x$ direction) Cubic fixed point and
towards the Heisenberg one. After it approaches this
attractive (in the $x$ direction)
fixed point the
evolution is similar to the one depicted in fig. 4.
Systems which start with larger values of $|\dkcr|$
behave similarly to fig. 7, but
deviate from the critical surface at earlier stages of the
evolution. As a result, they can feel the influence of
both the Cubic and Heisenberg fixed point, or only
the Cubic one, or they can deviate from the critical surface
too soon for any universal behaviour to be induced.
We calculate the renormalized parameters of the theory
(at various temperatures)
similarly to the previous subsection. Their behaviour as
a function of temperature
is analogous to that in fig. 5. The main
difference concerns the small region around $\tcr$.
In this region the temperature dependence should
reflect the
influence of the two fixed points during the evolution.
We first concentrate on values of $\dkcr$ for which the
critical behaviour is dominated by the cubic fixed point.
For this region we plot in
fig. 8 the critical exponents corresponding to
$\rhz(T)$ and $\xr(T)$, which are defined according to
\beq
\rhz(T) \propto& (\tcr^2 - T^2)^{2 \beta} \nonumber \\
2 - \xr(T) \propto& (\tcr^2-T^2)^{- \varphi}.
\label{nineeight} \eeq
We observe that they reach constant values as the critical
temperature is approached. The value for $\beta$ should
be characteristic of the Cubic fixed point.
We find
\be
\beta_C = 0.25
\label{ninenine} \ee
and $\nu_C = \zeta_C = 2 \beta_C$, in agreement with the
scaling laws and the finite value of the ratio
$\lr(T) T/m_R(T)$.
We expect the Cubic fixed point to generate the universal
behaviour characteristic of an Ising system. This is
due to the fact that the theory decomposes into
two disconnected
$Z_2$-symmetric theories for $x=2$ (see introduction
and sections 3 and 4). Indeed,
the critical exponents that we have calculated are
in exact agreement with the results of ref. \cite{transition}
for $N=1$,
which were obtained at the same level of the
truncation scheme.
Improved truncations result in values for the
exponents which are in agreement with
three-dimensional field theory \cite{amit,zinn}
at the few percent level \cite{peter}.
The exponent $\varphi=0.16$ is a typical example of a
crossover exponent \cite{aharony,amit}.
It is related to the growth of the unstable coupling
at the Cubic fixed point, and therefore to the
negative eigenvalue of the matrix which governs the
evolution of small perturbations around the
fixed point value of the parameters.
We postpone a more detailed discussion of the crossover
behaviour for a future publication \cite{peter}.

The behaviour shown in fig. 8 changes if
the critical temperature is further approached
(extension of the graph to the right).
Eventually the system moves away from the Cubic
fixed point and
the exponents $\beta,\nu,\zeta$ take values different
from those typical of an Ising system.
Also the temperature dependence of $2-\xr(T)$
cannot be described by a crossover exponent anymore and
$x_R(T)$ will rather follow eqs. (\ref{ninesix}),
(\ref{expmu}).
We display this behaviour in fig. 9. The values
of $\xr(T)$ give an indication of which fixed point
influences the system. It is
clear that the Heisenberg fixed point
takes over from the Cubic one very close to $\tcr$.
The temperature dependence of the exponent $\beta$
is a characteristic example of a crossover curve.
It demonstrates how the critical dynamics
changes from Ising-like (for $\ln \left(
\frac{T^2-\tcr^2}{\tcr^2} \right) \simeq -20$)
to Heisenberg-like (for $\ln
\left( \frac{T^2-\tcr^2}{\tcr^2} \right) \simeq -50$).
A detailed discussion of this behaviour within
more accurate truncation schemes will be
given in ref. \cite{peter}.

\setcounter{equation}{0}
\renewcommand{\theequation}{{\bf 11.}\arabic{equation}}

\section*{11. First order phase transition}

We turn now to the region $x > 2$ where we expect a first
order phase transition, as we have explained in sections
3 and 7. Our truncation scheme is too crude
to describe the behaviour of the potential in this region.
We have approximated $U_k(\rho_1,\rho_2,T)$
by a second order polynomial in $\rho_{1,2}$. This permits
the discussion of potentials with only one minimum.
The study of first order transitions requires the use
of improved truncations, where higher $\rho$-derivatives
of $U_k$ are taken into account and the possibility of
two distinct minima is permitted.
This has been done in ref. \cite{stefan} for the
zero temperature theory, and the existence of
a first order transition has been established.
We shall not repeat this calculation here. Instead we
shall derive an explicit solution of the evolution equation
in the region of large $x$, which will demonstrate the
existence of first order transitions for the high
temperature theory.

In fig. 10 we plot the numerical solution of eqs.
(\ref{sixsixteen})-(\ref{sixeighteen})
in the high temperature region, for
zero temperature renormalized parameters
$\lr=0.01$, $\xr=2.01,~3,~5$. The temperature is
very close to the critical one.
We notice that for all three sets of parameters the
evolution leads to a region of large $x$.
In fig. 10 the curves for $\kx$ and $\lx$ are
terminated when $x = 30$.
We observe that the running parameters
tend towards the same area of parameter space. More specifically,
for $x=30$ we find (very roughly)
$\lx \sim 3$, $\kx \sim 0.08$.
This convergence of flows was already apparent in fig. 1.
The difference in the evolution lies in the
``time'' $t = \ln(k/\Lx)$ that it takes
for the various flows to reach the same region.
The flows (a) and (b) are fast, while
the trajectory (c) starts very close to the
surface $x=2$, is first attracted towards the
Cubic fixed point, and finally deviates towards the
region of large $x$.
The Cubic fixed point separates the region
$x \le 2$, where we have observed second
order phase transitions, from the region $x > 2$,
for which we expect first order transitions.
For this reason
it is characterized as a tricritical point.
(The Ising fixed point exhibits similar behaviour.)

In the regions of large $x$
we have $\gb \gg \lb$.
As a result, the contribution of the $\phi_1$ fluctuations
to the evolution of
$U_k(\rhoa,0,T)$ is suppressed
as compared to the contribution of $\phi_2$.
Moreover, the increase of $x$ in this region is mainly due
to the fast decrease of $\lb = \lx k/T$.
In contrast, the coupling $\gb$
evolves only slowly. The $\rhoa$-dependent mass term for the
$\phi_2$ field is approximately given by
$\gb \rhoa$
(for $\rho_2 =0$ and
apart from a very small region around the origin)
and has again a mild $k$-dependence.
Let us assume that for a given scale $k_0$ the
solution of the
truncated evolution equations
depicted in fig. 10 gives a good approximation to the
exact solution for the potential.
(This means, in particular, that a two-minimum structure
has not appeared yet at this scale for the true potential.)
We denote the parameters of the theory
at the scale $k_0$ by
$\kx_0=\kx(k_0,T)$,
$\lb_0=\lb(k_0,T)$, $\gb_0=\gb(k_0,T)$,
$x_0=x(k_0,T)$, and the mass term for the
$\phi_2$ field by
$\gb_0 \rhoa$.
Based on the remarks at the beginning of this paragraph
we can obtain
in the high temperature region
an approximate solution of the
evolution equation (\ref{fourthree}) for the potential on
the $\rho_1$-axis ($\rho_2 =0$).
By neglecting the first term in the r.h.s. of eq. (\ref{fourthree})
and the $k$-dependence
of $U_2 = \frac{\partial U_k}{\partial \rhob}$,
the differential equation (\ref{fourthree}) is easily
integrated.
We obtain
in the limit $k \rightarrow 0$ (up to an irrelevant constant)
\beq
U(\rhoa,0,T) =&
U_{k=0}(\rhoa,0,T) \nonumber \\
=& \frac{1}{2} \lb_0 (\rhoa - \kx_0 k_0 T)^2
- \frac{T}{8 \pi^2} \int_0^{\infty} dx \sqrt{x}
\ln \left[ \frac{P_{k_0}(x) + \gb_0 \rhoa}{x+\gb_0 \rhoa} \right].
\label{eva} \eeq
The effective inverse propagator $P(x)$ is given by eq.
(\ref{twothirteen}) and we have indicated that it must be
evaluated for $k=k_0$.
Together with the numerical solution of the flow equations
near the critical surface for $k > k_0$, which provides
the ``integration constants''
$\lb_0$, $\gb_0$ and $\kx_0 k_0 T$,
we expect the effective potential of eq. (\ref{eva}) to be a
very good approximation. (For a sufficiently small ratio of couplings
$\lx_R/g_R$ we may identify $k_0$ with $T/\theta_2$. This essentially
reproduces the results of high temperature perturbation theory.)

The effective potential of eq. (\ref{eva}) describes indeed a first
order phase transition. This can be most easily visualized if we
approximate for the purpose of demonstration
\beq
P_{k_0} =& x~~~~~~~~~~~~~~~~~~~~~{\rm for}~~x>k_0^2 \nonumber \\
P_{k_0} =& k^2_0~~~~~~~~~~~~~~~~~~~~{\rm for}~~x<k_0^2.
\label{duo} \eeq
One finds for the $\rhoa$-derivative
\beq
U_1(\rhoa,0,T) =& \frac{\partial U(\rhoa,0,T)}{\partial \rhoa}
\nonumber \\
=& -\kx_0 \lb_0 k_0 T + \lb_0 \rhoa
+ \frac{\gb_0}{8 \pi^2} T \int_0^{k_0^2} dx \sqrt{x}
\left[ \frac{1}{x+\gb_0 \rhoa} - \frac{1}{k^2_0 + \gb_0 \rhoa}
\right].
\label{tria} \eeq
Using a rescaled field variable
\be
\rhb = \frac{\gb_0 \rhoa}{k^2_0}
\label{tessera} \ee
this yields (with $\lx_0 = \lb_0 T/k_0$)
\be
U_1(\rhb) = \frac{\gb_0 k_0 T}{4 \pi^2}
\biggl\{
\frac{2}{3} - \frac{4 \pi^2 \kx_0}{1+x_0}
-\sqrt{\rhb} \arctan \left( \frac{1}{\sqrt{\rhb}} \right)
+ \frac{4 \pi^2}{\lx_0 (1 +x_0)^2}\rhb
+\frac{1}{3} \frac{\rhb}{1 + \rhb}
\biggr\}.
\label{pevte} \ee
For $\kx_0 < \kx_A = \frac{1+x_0}{6 \pi^2}$ the potential
$U(\rhoa,0,T)$ develops a minimum at the origin
($\rhoa=0$). For $\kx_0$ only slightly below $\kx_A$
the origin is only a local minimum whereas the global minimum
occurs at $\rhoa \not= 0$ and the model is in the phase with
spontaneous symmetry breaking. For sufficiently small
$\kx_0/\kx_A$, however, the absolute minimum is at the origin
and the model is in the symmetric phase. (Note that
$\frac{2}{3}
-\sqrt{\rhb} \arctan \left( \frac{1}{\sqrt{\rhb}} \right)
+\frac{1}{3} \frac{\rhb}{1 + \rhb}$
is a positive function for all $\rhb$.)
There is a critical ratio
$\kx_0 / \kx_A$ (depending on the value of
$\lx_0 (1+x_0)^2$)
for which the minima at $\rhoa=0$ and $\rhoa \not= 0$
are degenerate in depth, but they are still well separated
from each other. Changing $\kx_0$ (which is a function of
$T$) through this critical value leads to a first order
phase transition with a jump in the order parameter.

The necessity of a first order phase transition can also be
seen by considering the $\rhoa$-dependent quartic coupling
$U_{11}(\rhoa) = \frac{\partial^2 U(\rhoa,0,T)}{\partial \rhoa^2}$
which obeys
\be
U_{11}(\rhoa) = \lb_0
-  \frac{\gb_0^2}{8 \pi^2} T \int_0^{\infty}
dx \sqrt{x} \left[
\frac{1}{(x+\gb_0 \rhoa)^2}
- \frac{1}{(P_{k_0}(x)+\gb_0 \rhoa)^2}
\right].
\label{nineten} \ee
By keeping only the most
singular behaviour of the integral for
$\rhoa \rightarrow 0$ we obtain
\be
U_{11}(\rhoa) = \lb_0 + \frac{\gb_0^2}{3 \pi^2} \frac{T}{k_0}
-  \frac{\gb_0^{3/2}}{16 \pi} T \frac{1}{\sqrt{\rhoa}}.
\label{ninetwelve} \ee
We have recovered the leading perturbative result for the
behaviour of the quartic coupling near a first order
phase transition.
If the minimum of the potential $\rho_{10}(k)$
is sufficiently close to zero at the scale $k_0$,
the remaining evolution
of $\rho_{10}(k)$
from $k_0$ to $k=0$ causes
$U_{11}$ to vanish at some scale $k$ between 0 and
$k_0$. As a consequence, the minimum at $\rho_{10} \not= 0$
becomes a saddlepoint and disappears subsequently.
Already before, a new minimum has been generated at the
origin, which remains the only minimum in the subsequent
evolution to $k =0$. Since $U_{11}(\rhoa)$ is always
negative for sufficiently small $\rhoa$, the phase transition can
never be second order and all values $x > 2$ must lead to a first
order phase transition.

Let us finally discuss a suitable choice of the scale $k_0$
from which on we can replace the numerical solution of the
flow equations (\ref{sixsixteen}) - (\ref{sixeighteen})
by the approximate solution given by eq. (\ref{eva}).
On one hand $x_0$ must be sufficiently large in order
to justify the neglection of the contribution of the
$\phi_1$-fluctuations in the approximate solution. On the
other hand $k_0$ should be sufficiently high so that
a second minimum at the origin has not yet been generated and
the truncation of a polynomial around $\rhz$ is still valid.
This requires that trajectories near the critical trajectory
not end at $k=0$ too deeply in the
symmetric phase. A realistic choice of $k_0$ should rather
correspond at $k=0$ to the situation where two minima exist
simultaneously.
For the
``quasicritical'' trajectories depicted in fig. 10
a reasonable compromise for $k_0$ seems to
be given by the value for which $x_0$ reaches 30.
(This corresponds to $\kx_A \simeq 0.6$.)
The trajectories (a), (b) and (c) shown in fig. 10 correspond
then to potentials $U(\rhoa,0,T)$ with two different minima,
as can be seen from fig. 11. They are close to, but not equal,
to the critical trajectories for which $\kx$
would deviate from fig. 10 towards the end of the running,
thus leading to a potential with two degenerate minima.

In summary, we have
established the occurence of a first order phase transition
for $x>2$. Moreover, we have reproduced the
perturbative prediction for the form of the potential
near the origin.
We should emphasize, however, that
the perturbative expression applies only to the
integration of fluctuations from the scale
$k_0$ (at which $x \gg 1$) to zero. The flow from the region
of $x$ near 2 to the region where the perturbative expression
becomes valid
can be computed only through the use of
evolution equations.
The different flows correspond to first order transitions
of varying strength. The discontinuity in the expectation
value is of the same order as
$k_0$. Also the mass gap at the critical temperature
is proportional to this scale. In consequence
the discontinuities in $\rho$
for the flows (a), (b), (c) in fig. 10
have a ratio of
$\Delta \rho_a / \Delta \rho_b / \Delta \rho_c
= 1/ 0.016 / 5.1 \times 10^{-9}$.
The last flow, which is remains in the vicinity
of the tricritical
point before deviating towards the region of large $x$,
corresponds to an extremely weakly first order
transition.

Our results can easily be extended to the region $x<0$.
We have seen in the introduction and section
4 that the AX regime ($x>0$) and the M regime
($x<0$) can be mapped onto each other
through a simple
redefinition of the fields (see eqs. (\ref{threeeleven}),
(\ref{threetwelve})).
For this reason, the physical behaviour in the two regimes
is the same. For example, the
Cubic and Ising fixed points generate the same
universal behaviour, characteristic of a $Z_2$-symmetric scalar theory.
Similarly, a first order phase transition occurs in the
region $x< -1$. We shall not repeat our discussion
for $x<0$. All our results can be
extended to this region by the
redefinition of fields and couplings according to
eqs. (\ref{threeeleven}), (\ref{threetwelve}).

\setcounter{equation}{0}
\renewcommand{\theequation}{{\bf 12.}\arabic{equation}}

\section*{12. Conclusions}

We have used the formalism of the effective average action
for the study of the high temperature phase transition
for a theory of two real scalar fields
$\chi_{1,2}$,
with the symmetry
$(\chi_1 \leftrightarrow - \chi_1,
\chi_2 \leftrightarrow - \chi_2,
\chi_1 \leftrightarrow  \chi_2)$,
and quartic potential
\be
V(\chi_1,\chi_2)
= \frac{1}{2} \mb (\chi_1^2+\chi_2^2)
+ \frac{1}{8} \lb (\chi_1^2+\chi_2^2)^2
+ \frac{1}{4} x \lb \chi_1^2\chi_2^2.
\label{tenone} \ee
The phase diagram of the theory is divided into
four disconnected regions:
$x>2$, $2>x>0$, $0>x>-1$, $x<-1$.
Three fixed points with at least one infrared stable
direction exist on the surfaces separating these regions:
The Heisenberg fixed point ($x=0$) is attractive in the
$\lx$ and $x$ directions, and corresponds to a theory
whose symmetry is increased to
$O(2)$.
The Cubic ($x=2$) and the Ising ($x=-1$)
fixed points are attractive in the $\lx$ direction and
repulsive in the $x$ direction and correspond
to two disconnected
$Z_2 (\chi_{1,2} \leftrightarrow - \chi_{1,2})$-symmetric
theories, which are equivalent.
The model has a second or first order phase transition, with
critical temperature well approximated  by the
perturbative expression if $\lb$ is small.

Theories with classical parameters in the regions
$2>x>0$, $0>x>-1$ have a second order phase transition. Very close
to the critical temperature the behaviour of the system
is universal.
It is characterized by critical exponents, which
are determined by the Heisenberg fixed point.
For theories with classical
parameters near the surfaces $x=2$, $x=-1$
the influence of the Cubic or Ising fixed point can
be observed near - but not too close to - the critical temperature.
This leads to a crossover phenomenon,
characterized by a crossover exponent and a crossover curve,
for temperatures approaching
$\tcr$. The universal behaviour is initially determined by
the Cubic or Ising fixed point for small enough $(T-\tcr)/\tcr$.
As the critical temperature is further
approached the more attractive Heisenberg fixed point dominates.
No part of this rich structure associated with the
second order phase transition can be observed
within perturbation theory.
We should mention that for small
values of $\lb$ the region in
temperature where these phenomena appear is rather narrow.
This changes for larger $\lb$, where the critical behaviour
extends over a larger temperature domain without
changing the universal results.
Even though we concentrated in the present paper on small values
of $\lb$ for the purpose of comparing with analytical results,
our method applies equally well to large $\lb$.

A first order phase transition is observed in the regions
$x>2$, $x<-1$. Therefore, the Cubic and Ising fixed points
are tricritical points separating regions
of second and first order transitions.
The perturbative expression for the effective potential is
a good approximation only for $x \gg 2$ and $x \simeq -2$.
All theories near the critical
temperature with classical couplings $x>2$ or $x<-1$
correspond to renormalized theories with
$x \gg 2$ or $x \simeq -2$ at scales of the order of the
mass gap of the model.
However, we distinguish two classes of theories:\\
I) For classical parameters $x \gg 2$ or $x \simeq -2$
one finds a strongly first order phase transition.
Here the effects of quantum or thermal fluctuations
are well approximated by the perturbative expression
for the effective potential.\\
II) For classical parameters $x \simeq 2$ or $x \simeq -1$
we predict a very weakly first order transition.
The use of the renormalization group is indispensable for
the correct incorporation of the quantum or thermal effects
which strongly renormalize the theory towards the regions
$x \gg 2$ or $x \simeq -2$.

Our results are relevant for multi-Higgs-scalar extensions
of the standard model
\cite{twoscalar} and multi-scalar
models of inflation \cite{inflation2}. They cast doubts
on the general validity of perturbative predictions
for the high temperature behaviour of these models
even in the case of small scalar couplings.
High temperature perturbation theory was
found to give a reliable estimate for the effective potential
only in limited regions of the parameter space.
Our non-perturbative method works for arbitrary values of the
couplings in eq. (\ref{tenone}). With straightforward modifications
- inclusion of $\chi^6$ couplings and anomalous dimension -
it gives quantitatively precise predictions for all temperatures
and all regions in the phase diagram. As demonstrated earlier
\cite{transition,indices} these modifications are sufficient
for computing critical exponents with a few percent accuracy.
The model is easily extended to the case where $\chi_1$ and
$\chi_2$ are $N$-component vectors with internal
$SO(N)$ symmetries. The high temperature phase
transition in other two-scalar models with a different
structure of the potential - as for example supersymmetric
two-doublet models - can be treated in complete analogy with the
present work.

\newpage

\setcounter{equation}{0}
\renewcommand{\theequation}{{\bf A.}\arabic{equation}}

\section*{Appendix A: The integrals $L^d_n$}

The integrals
$L^d_n(w)$ defined in eq. (\ref{threefive}), with the
choice of eqs. (\ref{twotwo}), (\ref{twothree})
for the infrared regulator, read
\be
L^d_n(\wt) = -2 n \int_0^{\infty} dy y^{\frac{d}{2}+1}
\frac{\exp(-y)}{\left[ 1-\exp(-y) \right]^2}
\left[ \frac{y}{1-\exp(-y)} + \wt \right]^{-(n+1)},
\label{aone} \ee
where $\wt = w/k^2$.
In the text we use $L^d_n(w)$ or $L^d_n(\wt)$ synonymously
for the integrals (\ref{aone}), even though they
depend only on the dimensionless ratio $\wt$. The integrals
obey the relations
\be
\frac{\partial}{\partial \wt} L^d_n(\wt) = -n L^d_{n+1}(\wt).
\label{aextra} \ee
For $\wt > -1$, the integrals
$|L^d_n(\wt)|$
are finite,
\footnote{
The pole structure of $L^d_n(\wt)$ is not relevant for this work.
For a discussion see ref. \cite{convex}.}
monotonically decreasing functions of $\wt$
We define
\be
L^d_n(0) = -2 l^d_n
\label{atwo} \ee
and give $l^d_n$ for the first three values of $n$
\beq
l^d_1 =~&\Gamma \left( \frac{d}{2} \right)
\nonumber \\
l^d_2 =~&2 \left( 1 - 2^{1-\frac{d}{2}} \right)
\Gamma \left( \frac{d}{2} - 1 \right)
\nonumber \\
l^d_3 =~&3 \left( 1 - 2^{3-\frac{d}{2}} + 3^{2-\frac{d}{2}} \right)
\Gamma \left( \frac{d}{2} - 2 \right).
\label{athree} \eeq
The asymptotic expansion of $L^d_n(\wt)$
for large arguments $\wt \rightarrow \infty$ is
\beq
L^d_n(\wt) =~&-2 n~ \zeta \left( \frac{d}{2} + 1 \right)
\Gamma \left( \frac{d}{2} + 2 \right) \wt^{-(n+1)}
\nonumber \\
&+ n (n+1) \left[
\zeta \left( \frac{d}{2} + 1 \right)
+ \zeta \left( \frac{d}{2} + 2 \right) \right]
\Gamma \left( \frac{d}{2} + 3 \right) \wt^{-(n+2)}...
\label{afive} \eeq

\newpage

\section*{Figures}

\renewcommand{\labelenumi}{Fig. \arabic{enumi}}
\begin{enumerate}
\item  
Flows on the $(\lx,x)$ plane for the three-dimensional
theory. The evolution is determined by eqs. (\ref{fiveseven}),
(\ref{fiveeight}) with $m^2=0$.
\vspace{6mm}
\item  
The integrals $L^3_1(w), L^3_2(w)$.
\vspace{6mm}
\item  
$L^4_1(w,T)/L^4_1(w)$ as a function of $T/k$,
for various values of $w/k^2$.
\vspace{6mm}
\item  
The evolution of $\kx$, $\lx$, $x$ in the high temperature region,
for temperatures slightly above and below the critical one, and
$\lr=0.01$, $\xr=1$, $\tcr^2/\rhz=4.78$.
The system approaches the Heisenberg fixed point
before deviating towards the symmetric phase or the phase
with spontaneous symmetry breaking.
\vspace{6mm}
\item  
The phase transition for a theory with
$\lr=0.01$, $\xr=1$, $\tcr^2/\rhz=4.78$.
\vspace{6mm}
\item  
The critical exponent $\beta(T)$ and the parameter $x_R(T)$
as the phase transition is
approached, for a theory with $\lx_R=0.2$, $x_R=1$.
\vspace{6mm}
\item  
The evolution of $\kx$, $\lx$, $x$ in the high temperature region,
for temperatures slightly above and below the critical one, and
$\lr=0.01$, $\xr$ slightly smaller than 2, $\tcr^2/\rhz=3.98$.
The system approaches first the Cubic and then the
Heisenberg fixed point,
before deviating towards the symmetric phase or the phase
with spontaneous symmetry breaking.
\vspace{6mm}
\item  
The critical exponents $\beta(T)$ and $\varphi(T)$
as the phase transition is
approached, for a theory with $\lx_R=0.5$ and $x_R$
slightly smaller than 2.
\vspace{6mm}
\item  
The critical exponent $\beta(T)$ and the parameter $x_R(T)$
as the phase transition is
approached, for a theory with $\lx_R=0.2$, $x_R=1.99$.
\vspace{6mm}
\item  
The evolution of $\kx$, $\lx$, $x$ in the high temperature region
for temperatures close to the critical ones.
The zero temperature parameters are: \\
a) $\lr=0.2$, $\xr=2.01$, \\
b) $\lr=0.2$, $\xr=3$, \\
c) $\lr=0.2$, $\xr=5$.
\item  
The effective potential ${\bar U} = {4 \pi^2 U}/{k^3_0 T} $
as a function of ${\bar \phi} = \sqrt{2 \gb_0 \rho_1}/{k_0}$,
resulting from the integration of eq. (\ref{pevte}) with
$\kx_0 = 0.08$, $\lx_0 = 3$ and $x_0 =30$.

\end{enumerate}

\end{document}